\documentclass[letterpaper,journal]{IEEEtran}

\usepackage{amsmath,amsfonts,amssymb}
\usepackage{algorithmic}
\usepackage{algorithm}
\usepackage{array}
\usepackage[caption=false,font=footnotesize]{subfig}
\usepackage{textcomp}
\usepackage{stfloats}
\usepackage{url}
\usepackage{verbatim}
\usepackage{graphicx}
\usepackage{cite}
\usepackage{bm}
\usepackage{booktabs}
\usepackage[colorlinks = true,
linkcolor = blue,
urlcolor  = blue,
citecolor = blue,
anchorcolor = blue]{hyperref}
\newcommand{\CN}{\mathcal{CN}}
\newcommand{\E}{\mathbb{E}}

\newcommand{\argmin}{\operatorname*{arg\,min}}
\newcommand{\argmax}{\operatorname*{arg\,max}}

\begin{document}

\title{Spatial-Code-Domain Grouped Index Modulation: Fluid-Antenna-Assisted System Design and BER Performance Analysis}

\author{Peng~Zhang,~\IEEEmembership{Graduate~Student~Member,~IEEE,} Jian~Dang,~\IEEEmembership{Senior~Member,~IEEE,} Yao~Ge,~\IEEEmembership{Member,~IEEE,} Miaowen~Wen,~\IEEEmembership{Senior~Member,~IEEE,} Ziyang~Liu, Liang~Wu,~\IEEEmembership{Senior~Member,~IEEE,} Zaichen~Zhang,~\IEEEmembership{Senior~Member,~IEEE,} and Yudong~Yao,~\IEEEmembership{Life~Fellow,~IEEE}

\thanks{
		

		
Peng~Zhang, Jian~Dang, Liang~Wu and Zaichen~Zhang are with the National Mobile Communications Research Laboratory, Frontiers Science Center for Mobile Information Communication and Security, Southeast University, Nanjing 211189, China; also with the Purple Mountain Laboratories, Nanjing 211111, China. (e-mail: peng\_zhang@seu.edu.cn, dangjian@seu.edu.cn, wuliang@seu.edu.cn, zczhang@seu.edu.cn).

Yao~Ge is with the AUMOVIO-NTU Corporate Laboratory, Nanyang Technological University, Singapore 639798 (e-mail: yao.ge@ntu.edu.sg).

Miaowen~Wen is with the School of Electronic and Information Engineering, South China University of Technology, Guangzhou 510640, China. (e-mail: eemwwen@scut.edu.cn).

Ziyang Liu is with the School of Communication Engineering, Hangzhou Dianzi University, Hangzhou 310018, China (e-mail: 251080010@hdu.edu.cn).

Yudong~Yao is with the Stevens Institute of Technology, Hoboken, NJ 07030
USA. (e-mail: yyao@stevens.edu).

Corresponding author: Jian~Dang and Zaichen~Zhang (dangjian@seu.edu.cn, zczhang@seu.edu.cn)
}}

\markboth{IEEE Transactions on Wireless Communications,~Vol.~XX, No.~XX, XXX~2026}
{Author \MakeLowercase{\textit{et al.}}: Fluid Antenna Assisted Code Grouped Index Modulation}

\maketitle

\begin{abstract}
Fluid antenna systems (FASs) provide reconfigurable spatial resources within compact apertures. In this paper, we introduce code-domain grouped index modulation (CGIM) and its spatial-code-domain extension, termed as SCGIM, for FA-assisted transceivers. CGIM partitions the available orthogonal spreading codes into multiple subsets and jointly maps information onto their in-phase and quadrature indices and constellation symbols. In an Rx-FAS-assisted single-input multiple-output (SIMO) link, group-wise despreading separates the orthogonal code groups for parallel detection, while receive-port selection provides spatial diversity. SCGIM further associates interleaved Tx-FA port subsets with the code subsets, with the Tx-FAS conveying spatial-index information and the Rx-FAS providing selection diversity in a multiple-input multiple-output (MIMO) link. For SCGIM, we develop maximum-likelihood (ML), staged greedy (GD), and cross-domain index message-passing (CD-IMPD) detectors. CD-IMPD exchanges soft information over a cycle-free factor graph to account for the coupling between the spatial and code indices, requiring only one inward and one outward message pass. For CGIM, the BER is derived from the joint decision regions of the despread-domain observations and averaged over the Rx-FAS selected-gain distribution under Rayleigh, Nakagami-\(m\), and additive white Gaussian noise channels. For SCGIM, an average-BER approximation is derived from a full-pair union bound using the selected-gain density ratio and exponentially tilted quadratic-form Laplace transforms. Simulation results validate the BER analysis and show that the proposed schemes achieve lower BER and higher throughput than the considered IM schemes, while CD-IMPD achieves near-ML BER performance with lower detection complexity.
\end{abstract}


\section{Introduction}
\subsection{Background}

Future wireless networks require high data rates and reliable transmission with limited spectrum, energy, and hardware resources~\cite{Tataria2021SixG}.
Conventional approaches based on high-order modulation, large antenna arrays,
or additional radio-frequency (RF) chains improve the data rate and
reliability at the cost of hardware, power, and signal-processing overhead.
Index modulation (IM) \cite{Zhang2026FARDIM} offers a complementary solution by conveying information
through both conventional symbols and the indices of transmission resources.
By activating selected antennas, subcarriers, time slots, or spreading codes,
IM enables flexible tradeoffs among spectral efficiency, energy efficiency,
and receiver complexity.

A fluid antenna system (FAS) provides multiple candidate ports within a compact
aperture and reconfigures its active radiating position
\cite{Wong2021FluidAntennaSystems}. At the receiver, port selection provides
spatial selection diversity. At the transmitter, a multi-RF-chain FAS can
simultaneously activate multiple ports
\cite{Zhu2024FAIM,Guo2025FluidAntennaIndexModulation}; when their activation
patterns are mapped to information bits, these ports serve as spatial-IM
entities. Thus, the receive FAS (Rx-FAS) mitigates deep fading, whereas the
transmit FAS (Tx-FAS) provides reconfigurable spatial resources for IM. Associating Tx-FA ports with orthogonal spreading codes introduces spatial-code indexing, while Rx-FA port selection provides receive diversity.

\subsection{Related Work}

Code index modulation (CIM) maps information bits onto spreading-code indices
and thereby conveys additional information through code selection under a
fixed transmit-energy normalization \cite{Kaddoum2015CIM}.
Generalized CIM (GCIM) jointly conveys the in-phase and quadrature components
of a modulation symbol through their spreading-code indices
\cite{Kaddoum2016GCIM}, while quadruple code index modulation (QCIM)
\cite{Cai2025QCIM} further extends
the number of code-index branches. GCIM has also been integrated with spatial modulation
\cite{Mesleh2008SM,Cogen2021GCIMSM}, while CIM-aided quadrature spatial
modulation \cite{Aydin2019CIMQSM} builds on the quadrature spatial modulation
principle in \cite{Mesleh2015QSM}. However, these schemes use predefined code-index structures and do not consider grouping multiple orthogonal code branches for scalable transmission and detection.

FA-assisted IM initially employed a single activated FA port to convey index
bits along with conventional symbols, with channel-coded port patterns further
improving robustness against spatial correlation
\cite{Chen2024FAIM,Faddoul2025CodedFAIM}. This principle was subsequently
extended to multiple-port activation for FA-assisted multiple-input
multiple-output (MIMO) transmission \cite{Zhu2024FAIM}. Grouped FA-IM
subsequently partitioned the FA ports into spatially structured sets to support
parallel transmission and low-complexity detection
\cite{Guo2025FluidAntennaIndexModulation}. Existing schemes
nevertheless focus primarily on the FA port itself as the indexed resource.
However, joint mapping and detection across the FA spatial and orthogonal code domains have received limited attention.

Fundamental studies of FAS quantified its performance limits under correlated Rayleigh
fading \cite{Wong2020PerformanceLimits} and established outage, capacity, and
second-order statistics for port selection \cite{Wong2021FluidAntennaSystems}.
Closed-form reference-port correlation parameters were derived in
\cite{Wong2022CorrelationParameters}, followed by a refined characterization
of outage probability and diversity gain \cite{New2024FASOutage}. Spatial
block-correlation \cite{RamirezEspinosa2024BlockCorrelation} and Gaussian
copula models \cite{Ghadi2024GaussianCopula} provided tractable alternatives
for modeling spatial dependence. The analysis was then
extended to Nakagami-$m$ \cite{VegaSanchez2024SimpleMethod},
$\alpha$-$\mu$ \cite{Alvim2023AlphaMu}, and Rician fading
\cite{Huangfu2026Rician}. Subsequent studies derived error probabilities for
conventional modulation \cite{Zhu2026GeometricSER} and block error rates under
finite-blocklength transmission \cite{Zhang2025UniversalBLER}. These studies show that FAS reliability depends on the aperture, spatial correlation, and number of candidate ports. Nevertheless, existing error
analyses mainly address constellation decisions or specific single-domain FA-IM schemes and
do not directly characterize the coupled errors among Tx-port indices, code
indices, constellation symbols, and correlated Rx-FA selection.

\subsection{Motivation and Contributions}

Grouping and block-wise IM decompose high-dimensional index spaces into parallel
low-dimensional subspaces. With preserved subcarrier orthogonality,
frequency-domain groups separate exactly \cite{Basar2013OFDMIM}, whereas
delay--Doppler- and affine-frequency-domain mappings reduce the search dimension
but generally retain inter-block coupling
\cite{Qian2023BlockwiseOTFSIM,Zhu2024AFDMIM,Qian2025GCIMAFDM}. Similar grouping has also been
applied to reconfigurable intelligent surface elements and FA ports
\cite{Jin2023RISGrouping,Guo2025FluidAntennaIndexModulation,Weng2026RISGroupedFAS}.
Orthogonal spreading codes can also be separated exactly by despreading. This property allows the code domain to be organized into multiple groups.

Existing FA-assisted IM studies mainly consider an FAS at either the transmitter or the receiver, while joint Tx- and Rx-FAS deployment has received limited attention. Such a system requires joint mapping and detection of the Tx-port indices, code indices, and constellation symbols together with Rx-FA port selection. Its BER analysis must further account for Tx-port correlation in the spatial-code pairwise distances, Rx-port selection statistics, and bit-label weights. Therefore, conventional IM and single-sided FAS analyses cannot be directly applied

To this end, we develop an orthogonal-code-domain grouped IM framework for
FA-assisted transceivers. It first establishes grouped code-domain transmission
for an Rx-FAS-assisted single-input multiple-output (SIMO) link and then
couples orthogonal code groups with Tx-FA port groups for joint
spatial-code-domain IM. The main contributions are summarized as follows.

\begin{itemize}

\item We propose code-domain grouped index modulation (CGIM), which partitions
the spreading codes into orthogonal groups and jointly maps their in-phase and
quadrature indices with constellation symbols. Group-wise despreading separates
the orthogonal code groups and enables parallel detection. Under a fixed spreading length and quadrature amplitude
modulation (QAM) order, the additional code-index dimensions can increase the
spectral efficiency.

\item We further extend CGIM to spatial-code-domain grouped index modulation (SCGIM)
for an FA-assisted MIMO transceiver. SCGIM associates interleaved Tx-FA port
groups with orthogonal code groups to support Tx-FA spatial-index transmission
and Rx-FA selection diversity.

\item We develop maximum-likelihood (ML), staged greedy detection (GD), and
cross-domain index message-passing detection (CD-IMPD) for SCGIM. On the resulting cycle-free factor graph, CD-IMPD preserves the spatial-code coupling through soft-information exchange, while a single inward and outward message-passing sweep yields the exact variable-node marginals under the factorized likelihood model.

\item We derive the BER performance of both schemes. A joint decision-region
analysis yields analytical or semi-analytical CGIM expressions under Rayleigh,
Nakagami-$m$, and additive white Gaussian noise (AWGN) channels. or SCGIM, a
full-pair union bound is combined with the Rx-FAS selected-gain density ratio
and exponentially tilted quadratic-form Laplace transforms to derive an
average BER approximation.

\end{itemize}

The rest of this paper is organized as follows. Section~\ref{sec:system_model}
presents the correlated Tx- and Rx-FAS channel models.
Sections~\ref{sec:cgim_design} and \ref{sec:scgim_design} develop the CGIM and
SCGIM transceivers, respectively,
including their mapping and detection schemes. Section~\ref{sec:se_complexity}
analyzes the spectral efficiency and detection complexity, and
Section~\ref{sec:theoretical_analysis} derives the BER performance of both
schemes. Section~\ref{sec:simulations} presents the simulation results and
discussion. Finally, Section~\ref{sec:conclusion} concludes this paper.

\emph{Notations:}
Bold lowercase and uppercase letters denote vectors and matrices,
respectively. The sets of binary, real, and complex numbers are denoted by
$\mathbb B=\{0,1\}$, $\mathbb R$, and $\mathbb C$, respectively, while calligraphic
letters denote finite sets, codebooks, or constellations; $|\mathcal A|$ is
the cardinality of $\mathcal A$. The imaginary unit is
$\mathrm j=\sqrt{-1}$, and $\mathbf I_n$ denotes the $n\times n$ identity
matrix. For a complex scalar $x$, $|x|$ is its magnitude, and $\Re\{\cdot\}$ and
$\Im\{\cdot\}$ extract the real and imaginary parts. The operators
$(\cdot)^*$, $(\cdot)^{\operatorname T}$, and
$(\cdot)^{\operatorname H}$ denote complex conjugation, transpose, and
conjugate transpose, respectively. The Euclidean and Frobenius norms are
$\|\cdot\|$ and $\|\cdot\|_{\operatorname F}$, respectively.
The operators $\operatorname{tr}(\cdot)$, $\det(\cdot)$,
$\operatorname{rank}(\cdot)$, $\operatorname{vec}(\cdot)$, and
$\lambda_{\max}(\cdot)$ denote the trace, determinant, rank, vectorization,
and largest eigenvalue, respectively. The operators $\argmin$ and $\argmax$
return the minimizing and maximizing arguments, respectively, and
$\lfloor\cdot\rfloor$ is the floor operator. The expectation, probability, and
variance operators are $\mathbb E[\cdot]$, $\Pr\{\cdot\}$, and
$\operatorname{Var}(\cdot)$; $f_X(\cdot)$, $F_X(\cdot)$, and
$\mathcal L_X(\cdot)$ denote the probability density function (PDF),
cumulative distribution function (CDF), and Laplace transform of a random
variable $X$, respectively. The Gaussian $Q$-function is denoted by
$Q(\cdot)$, and a circularly symmetric complex Gaussian random vector follows
$\mathcal{CN}(\boldsymbol\mu,\mathbf C)$.

\section{Fluid Antenna System and Channel Model}
\label{sec:system_model}

We consider a block-fading wireless link whose receiver employs an Rx-FAS.
Its candidate ports are partitioned into $N_{\mathrm r}$ disjoint groups,
with one RF chain assigned to each group. The $r$-th group contains $K_r$
candidate ports distributed over a local aperture of length
$W_{\mathrm r}\lambda$. Ports within the same group experience spatially
correlated fading, whereas different groups are sufficiently separated and
are modeled as statistically independent.

Two transmitter configurations are considered. CGIM employs a single
fixed-position antenna and forms an Rx-FAS-assisted SIMO link, whereas SCGIM
employs a Tx-FAS with $K_{\mathrm t}$ candidate ports and $G$ RF chains and
forms an FA-assisted MIMO link. Their mapping and receiver designs are
presented in Sections~\ref{sec:cgim_design} and \ref{sec:scgim_design},
respectively. Each transmission block is shorter than the channel coherence
time, such that all channel coefficients remain constant within one block.

For the transmit side, the single fixed antenna employed by CGIM does not
involve transmit-port selection or transmit-side spatial correlation. In the
SCGIM configuration, the $K_{\mathrm t}$ candidate ports of the Tx-FAS are
uniformly distributed over an aperture of length $W_{\mathrm t}\lambda$.
For $K_{\mathrm t}>1$, the normalized displacement of transmit port $k$
from the reference port is
$\zeta_k=(k-1)W_{\mathrm t}/(K_{\mathrm t}-1)$, and the corresponding
reference correlation coefficient is
$\nu_k=J_0(2\pi\zeta_k)$, where $J_0(\cdot)$ denotes the zeroth-order
Bessel function of the first kind and $\nu_1=1$. The Tx-FAS correlation
matrix $\mathbf R_{\mathrm t}\in
\mathbb C^{K_{\mathrm t}\times K_{\mathrm t}}$ has entries
\begin{equation}
    [\mathbf R_{\mathrm t}]_{i,j}
    =
    \begin{cases}
        1, & i=j,\\
        \nu_i\nu_j, & i\ne j.
    \end{cases}
    \label{eq:tx_corr_entries}
\end{equation}
For $K_{\mathrm t}=1$, the Tx-FAS reduces to a fixed-position antenna and
$\mathbf R_{\mathrm t}=1$.

At the receiver, let $h_{r,q}$ denote the scalar channel from the fixed
transmit antenna to the $q$-th candidate port in receive-port group $r$,
where $r\in\{1,\ldots,N_{\mathrm r}\}$ and
$q\in\{1,\ldots,K_r\}$. Taking the first port as the reference, the
correlated channels are represented as
\begin{equation}
    h_{r,q}
    =\mu_q h_{r,1}
    +\sqrt{1-\mu_q^2}\,e_{r,q},
    \qquad q=2,\ldots,K_r,
    \label{eq:rx_reference_model}
\end{equation}
where $h_{r,1}\sim\CN(0,1)$ and $e_{r,q}\sim\CN(0,1)$. The variables
$\{e_{r,q}\}$ are mutually independent and are also independent of the
reference channel. The reference correlation coefficient is
\begin{equation}
    \mu_q=J_0(2\pi\xi_q),
    \qquad q=1,\ldots,K_r,
    \label{eq:rx_reference_coefficient}
\end{equation}
where $\xi_q=(q-1)W_{\mathrm r}/(K_r-1)$ is the normalized displacement
of port $q$ from the reference port for $K_r>1$. When $K_r=1$, the
corresponding receive-port group reduces to a fixed-position antenna with
$\mu_1=1$.

Define
$\mathbf h_r=[h_{r,1},h_{r,2},\ldots,h_{r,K_r}]^{\operatorname T}$.
Its receive-side correlation matrix is
$\mathbf J_r=\E\{\mathbf h_r\mathbf h_r^{\operatorname H}\}$, with
\begin{equation}
    [\mathbf J_r]_{p,q}
    =
    \begin{cases}
        1, & p=q,\\
        \mu_p\mu_q, & p\ne q.
    \end{cases}
    \label{eq:rx_corr_entries}
\end{equation}

For the MIMO configuration, define the channel-response vector associated
with candidate port $q$ in receive-port group $r$ as
$\mathbf h_{r,q}=[h_{r,q,1},\ldots,h_{r,q,K_{\mathrm t}}]
\in\mathbb C^{1\times K_{\mathrm t}}$, where $h_{r,q,k}$ is the channel
coefficient from transmit port $k$ to candidate receive port $q$. The
candidate-port channel matrix is
$\mathbf H_r=[\mathbf h_{r,1}^{\operatorname T},\ldots,
\mathbf h_{r,K_r}^{\operatorname T}]^{\operatorname T}
\in\mathbb C^{K_r\times K_{\mathrm t}}$. Combining the transmit- and
receive-side correlation models, $\mathbf H_r$ is modeled as
\begin{equation}
    \mathbf H_r
    =\mathbf J_r^{1/2}
     \mathbf H_{\mathrm w,r}
     \mathbf R_{\mathrm t}^{1/2},
    \label{eq:mimo_channel_model}
\end{equation}
where
$\operatorname{vec}(\mathbf H_{\mathrm w,r})
\sim\CN(\mathbf 0,\mathbf I_{K_rK_{\mathrm t}})$.
The matrices $\{\mathbf H_r\}_{r=1}^{N_{\mathrm r}}$ are mutually
independent owing to the assumed separation between the receive-port groups.

After each channel realization, the Rx-FAS selects one port from each
receive-port group. For the SIMO configuration, the $r$-th RF chain selects
the port with the largest instantaneous channel power:
\begin{equation}
    q_r^\star
    =\argmax_{q\in\{1,\ldots,K_r\}}|h_{r,q}|^2.
    \label{eq:simo_rx_selection}
\end{equation}
The equivalent SIMO channel is
$\mathbf h_{\mathrm{sel}}=[h_{1,q_1^\star},\ldots,
h_{N_{\mathrm r},q_{N_{\mathrm r}}^\star}]^{\operatorname T}
\in\mathbb C^{N_{\mathrm r}\times1}$.

For the MIMO configuration, each candidate receive port is characterized by
its channel-response vector to all Tx-FAS ports. Accordingly, the $r$-th RF
chain selects the candidate port with the largest instantaneous channel
energy:
\begin{equation}
    q_r^\star
    =\argmax_{q\in\{1,\ldots,K_r\}}
      \|\mathbf h_{r,q}\|^2.
    \label{eq:mimo_rx_selection}
\end{equation}
The selected channel response of receive-port group $r$ is
$\mathbf h_r^\star=\mathbf h_{r,q_r^\star}$. The equivalent MIMO channel is
$\mathbf H_{\mathrm{sel}}
=[(\mathbf h_1^\star)^{\operatorname T},\ldots,
(\mathbf h_{N_{\mathrm r}}^\star)^{\operatorname T}]^{\operatorname T}
\in\mathbb C^{N_{\mathrm r}\times K_{\mathrm t}}$.

\section{CGIM Transceiver Design}
\label{sec:cgim_design}

\begin{figure*}[!t]
    \centering
    \includegraphics[width=0.7\textwidth]{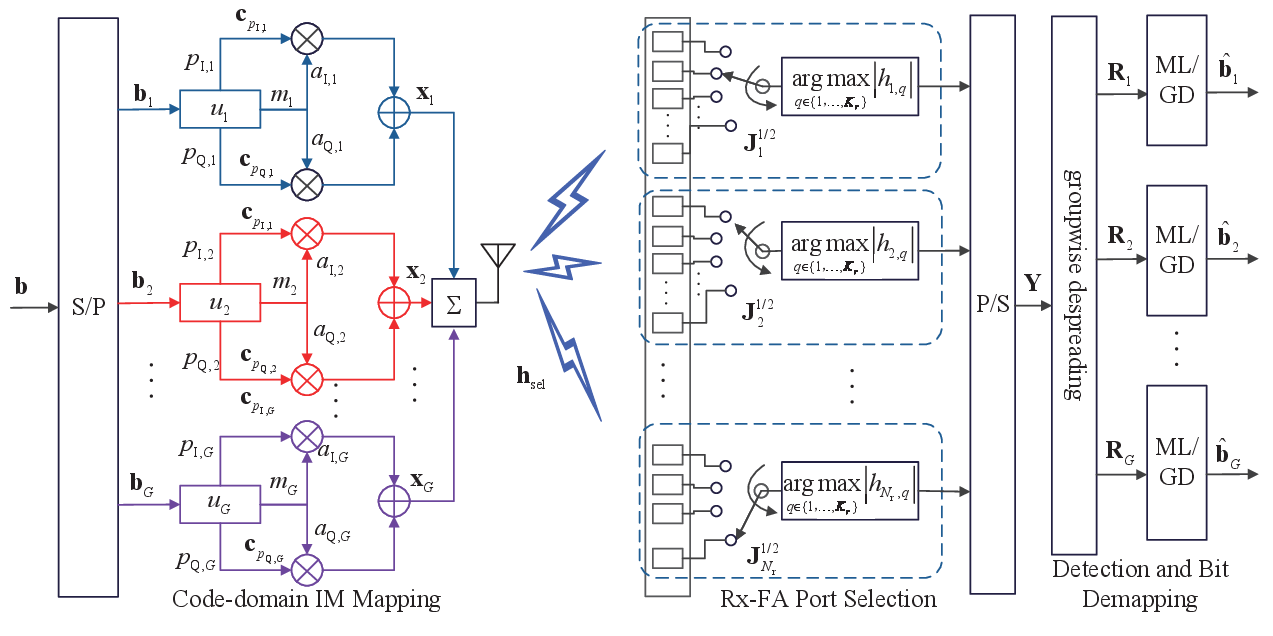}
    \vspace{-0.08in}
    \caption{System block diagram of the proposed CGIM scheme.}
    \label{fig:cgim_system_model}
    \vspace{-0.5cm}
\end{figure*}

\subsection{Code-Domain Grouping and Index Mapping}
As shown in Fig.~\ref{fig:cgim_system_model}, CGIM employs a single fixed
transmit antenna and a bank of length-$L$ orthogonal Walsh spreading codes.
From this bank, $F=GP\leq L$ codes are assigned to $G$ disjoint codebooks,
each containing $P$ candidates. Let
$\mathcal P_g=\{p_{g,1},p_{g,2},\ldots,p_{g,P}\}$ denote the code-index set
of group $g$, so that the $g$-th codebook is
$\mathcal C_g=\{\mathbf c_p:p\in\mathcal P_g\}$. Specifically, a binary
sequence $\mathbf d_p\in\mathbb B^{1\times L}$ is mapped to the normalized bipolar
spreading code $\mathbf c_p=(2\mathbf d_p-\mathbf 1_L^{\operatorname{T}})/\sqrt L$.
Then $\mathbf c_p\in
\{-1/\sqrt L,+1/\sqrt L\}^{1\times L}$, and the normalized codes satisfy
\begin{equation}
    \mathbf c_p\mathbf c_{p'}^{\operatorname{T}}
    =
    \begin{cases}
        1, & p=p',\\
        0, & \text{otherwise}.
    \end{cases}
    \label{eq:cgim_code_orthogonality}
\end{equation}

Let $\mathbf b=[\mathbf b_1,\ldots,\mathbf b_G]$ be the input bit vector of
one transmission block. For group $g$,
$\mathbf b_g=[\mathbf b_{\mathrm{I},c,g},\mathbf b_{\mathrm{Q},c,g},\mathbf b_{\mathrm{s},g}]$
contains two $\log_2P$-bit code-index labels and one $\log_2M$-bit
modulation label. The first two labels determine
$p_{\mathrm{I},g},p_{\mathrm{Q},g}\in\mathcal P_g$, while $\mathbf b_{\mathrm{s},g}$ selects an
$M$-ary quadrature amplitude modulation ($M$-QAM) symbol
$s_g\in\mathcal S_M$. Equivalently, the group mapping
produces $u_g=(p_{\mathrm{I},g},p_{\mathrm{Q},g},m_g)\in\mathcal U_g^{\mathrm{CGIM}}$, where
$\mathcal U_g^{\mathrm{CGIM}}=\mathcal P_g^2\times\{1,\ldots,M\}$.
One group conveys
$B_{g}^{\mathrm{CGIM}}=2\log_2P+\log_2M$ bits, and one block conveys
$B^{\mathrm{CGIM}}=GB_{g}^{\mathrm{CGIM}}$ bits.

For square $M$-QAM, write $s_g=a_{\mathrm{I},g}+\mathrm j a_{\mathrm{Q},g}$, where
$a_{\mathrm{I},g},a_{\mathrm{Q},g}\in\mathcal A$ and $\mathcal A$ is the corresponding
$\sqrt{M}$-ary pulse amplitude modulation (PAM) alphabet. The I- and Q-components independently select their
spreading codes, and the two selections may coincide; when
$p_{\mathrm{I},g}=p_{\mathrm{Q},g}$, the same code conveys both components.

For the mapped tuple $u_g=(p_{\mathrm{I},g},p_{\mathrm{Q},g},m_g)$, the chip vector
$\mathbf x_g(u_g)\in\mathbb C^{1\times L}$ of group $g$ is
\begin{equation}
    \mathbf x_g(u_g)
    =\frac{1}{\sqrt G}\left(
      a_{\mathrm{I},g}\mathbf c_{p_{\mathrm{I},g}}
      +\mathrm j a_{\mathrm{Q},g}\mathbf c_{p_{\mathrm{Q},g}}\right).
    \label{eq:cgim_group_signal}
\end{equation}
The transmitted chip vector $\mathbf x(\mathbf u)\in\mathbb C^{1\times L}$
is formed by superimposing the $G$ group signals as
\begin{equation}
    \mathbf x(\mathbf u)=\sum_{g=1}^{G}\mathbf x_g(u_g)
    =\frac{1}{\sqrt G}\sum_{g=1}^{G}
      \left(a_{\mathrm{I},g}\mathbf c_{p_{\mathrm{I},g}}
      +\mathrm j a_{\mathrm{Q},g}\mathbf c_{p_{\mathrm{Q},g}}\right),
    \label{eq:cgim_transmit_signal}
\end{equation}
where $\mathbf u=(u_1,\ldots,u_G)$. The factor $1/\sqrt G$ keeps the
average block energy invariant with $G$. Under this formulation, conventional
GCIM and QCIM correspond to $G=1$ and $G=2$, respectively, whereas CGIM
generalizes the number of code groups.

\subsection{CGIM Receiver Design}
After Rx-FAS port selection, the received block
$\mathbf Y\in\mathbb C^{N_{\mathrm r}\times L}$ is
\begin{equation}
    \mathbf Y=\mathbf h_{\mathrm{sel}}\mathbf x(\mathbf u)+\mathbf N,
    \label{eq:cgim_received_signal}
\end{equation}
where $\mathbf N\in\mathbb C^{N_{\mathrm r}\times L}$ contains independent
$\CN(0,N_0)$ entries. Let $E_{\mathrm{s}}$ denote the average energy of one CGIM block
and $\gamma_{\mathrm{b}}=E_{\mathrm{b}}/N_0$. The noise variance is
$N_0=E_{\mathrm{s}}/(B^{\mathrm{CGIM}}\gamma_{\mathrm{b}})$.

The spreading-code matrix of group $g$ is defined as
$\mathbf C_g=[\mathbf c_{p_{g,1}}^{\operatorname{T}},\ldots,\mathbf c_{p_{g,P}}^{\operatorname{T}}]^{\operatorname{T}}
\in\mathbb R^{P\times L}$. The receiver performs group-wise despreading on the
$N_{\mathrm r}$ observations using all codes in $\mathcal C_g$, yielding the
despread-domain observation matrix
\begin{equation}
    \mathbf R_g
    =\mathbf Y\mathbf C_g^{\operatorname{T}}
    =[\mathbf r_{g,1},\ldots,\mathbf r_{g,P}]
    \in\mathbb C^{N_{\mathrm r}\times P},
    \label{eq:cgim_groupwise_despreading}
\end{equation}
where $\mathbf r_{g,p}=\mathbf Y\mathbf c_{p_{g,p}}^{\operatorname{T}}\in
\mathbb C^{N_{\mathrm r}\times1}$ denotes the despread observation vector
associated with code label $p_{g,p}$. By \eqref{eq:cgim_code_orthogonality}, its input--output
relation is
\begin{equation}
\mathbf r_{g,p}
=\frac{\mathbf h_{\mathrm{sel}}}{\sqrt G}
\left[
a_{\mathrm{I},g}\mathbf c_{p_{\mathrm{I},g}}
\mathbf c_{p_{g,p}}^{\operatorname T}
+\mathrm j a_{\mathrm{Q},g}\mathbf c_{p_{\mathrm{Q},g}}
\mathbf c_{p_{g,p}}^{\operatorname T}
\right]
+\widetilde{\mathbf n}_{g,p},
\label{eq:cgim_despread_output}
\end{equation}
where
$\widetilde{\mathbf n}_{g,p}=\mathbf N\mathbf c_{p_{g,p}}^{\operatorname{T}}
\sim\CN(\mathbf0,N_0\mathbf I_{N_{\mathrm r}})$. By code orthogonality, the
I- and Q-components are confined to the columns indexed by
$p_{\mathrm{I},g}$ and $p_{\mathrm{Q},g}$, respectively; if the two indices
coincide, both components occupy the same despread vector.

The ML detector jointly recovers the two code indices and the QAM label.
For a legal candidate $u=(p_{\mathrm{I}},p_{\mathrm{Q}},m)\in\mathcal U_g^{\mathrm{CGIM}}$, write
$s_m=a_{\mathrm{I},m}+\mathrm j a_{\mathrm{Q},m}$ and let $\mathbf e_p\in\mathbb R^{P\times1}$ denote
the $p$-th standard basis vector. The corresponding noiseless despread-domain response is
$\overline{\mathbf R}_g(u)=\mathbf h_{\mathrm{sel}}
(a_{\mathrm{I},m}\mathbf e_{p_{\mathrm{I}}}^{\operatorname{T}}+\mathrm j a_{\mathrm{Q},m}\mathbf e_{p_{\mathrm{Q}}}^{\operatorname{T}})/\sqrt G$.
The group-wise ML decision is
\vspace{-0.2cm}
\begin{equation}
\left\langle\widehat p_{\mathrm{I},g},\widehat p_{\mathrm{Q},g},\widehat m_g\right\rangle
=\argmin_{u\in\mathcal U_g^{\mathrm{CGIM}}}
\left\|\mathbf R_g-\overline{\mathbf R}_g(u)\right\|_{\operatorname{F}}^2.
\label{eq:cgim_joint_ml_detector}
\vspace{-0.1cm}
\end{equation}
To expose the sufficient statistics of the despread-domain ML search, define
$z_{g,p}=\mathbf h_{\mathrm{sel}}^{\operatorname{H}}\mathbf r_{g,p}
=\sum_{r=1}^{N_{\mathrm r}}h_r^*r_{g,p}^{(r)}$, which coherently combines the $N_{\mathrm r}$
despread observations associated with code $p$. Expanding the squared distance
in \eqref{eq:cgim_joint_ml_detector} gives
$\|\mathbf R_g-\overline{\mathbf R}_g(u)\|_{\operatorname{F}}^2
=\|\mathbf R_g\|_{\operatorname{F}}^2+D_g(u)$, where
\begin{equation}
D_g(u)=-\frac{2}{\sqrt G}\left[
a_{\mathrm{I},m}\Re\{z_{g,p_{\mathrm{I}}}\}
+a_{\mathrm{Q},m}\Im\{z_{g,p_{\mathrm{Q}}}\}\right]
+\frac{\|\mathbf h_{\mathrm{sel}}\|^2|s_m|^2}{G}.
\label{eq:cgim_expanded_ml_metric}
\end{equation}
Removing the candidate-independent term
$\|\mathbf R_g\|_{\operatorname{F}}^2$, ML reduces to minimizing
\eqref{eq:cgim_expanded_ml_metric} over the $P^2M$ candidates of each group.
Code orthogonality decouples the group metrics, so independent group-wise
minimization yields the blockwise ML decision. Each detected tuple is then
demapped to $\widehat{\mathbf b}_g$.

\emph{Greedy detection:}
For square $M$-QAM, code orthogonality and the Cartesian PAM structure make
the metric in \eqref{eq:cgim_expanded_ml_metric} exactly separable across the
I/Q branches. The resulting ML-equivalent greedy detector (GD) first performs
channel-matched combining on every despread vector. Since the factors $\sqrt G$ and
$\|\mathbf h_{\mathrm{sel}}\|^{-2}$ are common to all candidate codes, the
active I/Q code indices are selected directly from the magnitudes of the
corresponding matched-filter outputs as
\begin{subequations}\label{eq:cgim_gd_index_detection}
\begin{align}
\widehat p_{\mathrm{I},g}&=\arg\max_p
\left|\Re\{\mathbf h_{\mathrm{sel}}^{\operatorname{H}}\mathbf r_{g,p}\}\right|,
\label{eq:cgim_gd_index_detection_I}\\
\widehat p_{\mathrm{Q},g}&=\arg\max_p
\left|\Im\{\mathbf h_{\mathrm{sel}}^{\operatorname{H}}\mathbf r_{g,p}\}\right|.
\label{eq:cgim_gd_index_detection_Q}
\end{align}
\end{subequations}
Conditioned on these indices, the normalized matched-filter outputs are sliced
to the nearest PAM levels, after which the corresponding QAM label is obtained
as
\begin{subequations}\label{eq:cgim_gd_qam_decision}
\begin{align}
\widehat a_{\mathrm{I},g}&=\arg\min_{a\in\mathcal A}
\left|\frac{\sqrt G\Re\{\mathbf h_{\mathrm{sel}}^{\operatorname{H}}
\mathbf r_{g,\widehat p_{\mathrm{I},g}}\}}
{\|\mathbf h_{\mathrm{sel}}\|^2}-a\right|,
\label{eq:cgim_gd_qam_decision_I}\\
\widehat a_{\mathrm{Q},g}&=\arg\min_{a\in\mathcal A}
\left|\frac{\sqrt G\Im\{\mathbf h_{\mathrm{sel}}^{\operatorname{H}}
\mathbf r_{g,\widehat p_{\mathrm{Q},g}}\}}
{\|\mathbf h_{\mathrm{sel}}\|^2}-a\right|,
\label{eq:cgim_gd_qam_decision_Q}\\
\widehat m_g&=\arg\min_{m\in\{1,\ldots,M\}}
|s_m-(\widehat a_{\mathrm{I},g}+\mathrm j\widehat a_{\mathrm{Q},g})|^2.
\label{eq:cgim_gd_qam_decision_symbol}
\end{align}
\end{subequations}
Finally, $(\widehat p_{\mathrm{I},g},\widehat p_{\mathrm{Q},g},\widehat m_g)$ is inverse-mapped
to the detected bits.
\vspace{-0.6cm}
\section{SCGIM Transceiver Design}
\label{sec:scgim_design}

\begin{figure*}[!t]
    \centering
    \includegraphics[width=0.8\textwidth]{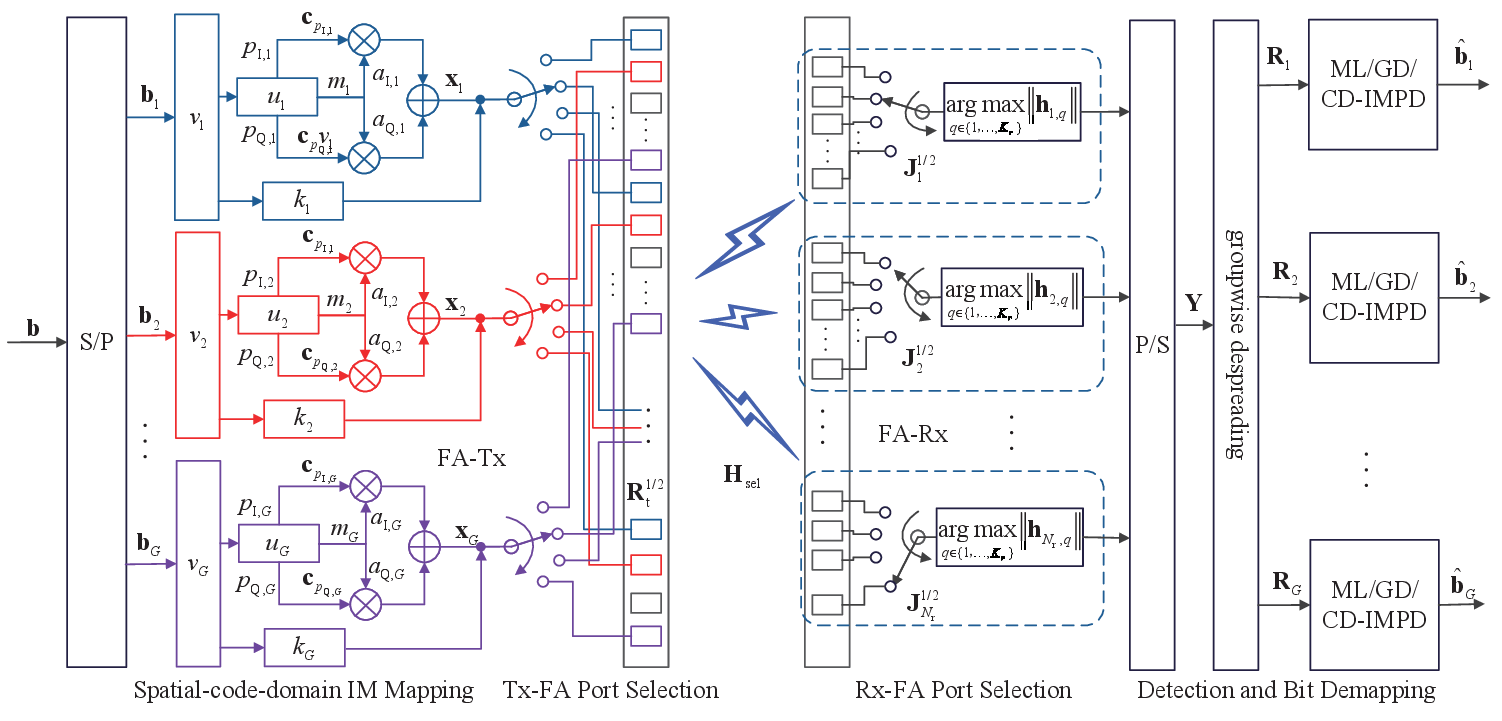}
    \vspace{-0.16in}
    \caption{System block diagram of the proposed SCGIM scheme.}
    \label{fig:scgim_system_model}
    \vspace{-0.5cm}
\end{figure*}

\subsection{Interleaved Tx-Port Grouping and Index Mapping}
As illustrated in Fig.~\ref{fig:scgim_system_model}, SCGIM employs a Tx-FAS
with $K_{\mathrm{t}}$ physical ports and $G$ RF chains. Assuming that
$K_{\mathrm{t}}$ is divisible by $G$, each RF chain is associated with
$K_{\mathrm{t},g}=K_{\mathrm{t}}/G$ candidate ports and one code-domain group.
To distribute the ports of each group across the Tx-FAS aperture, the $k$-th
candidate of group
$g$ is mapped to the physical port
$\pi_g(k)=g+(k-1)G$, where $k\in\{1,\ldots,K_{\mathrm{t},g}\}$. Hence, the interleaved
port set is $\mathcal S_g=\{g,g+G,\ldots,g+(K_{\mathrm{t},g}-1)G\}$. Compared with a
contiguous partition, this mapping increases the separation between adjacent
candidates within the same group while preserving disjoint port sets across
the $G$ RF chains.

Let $u_g=(p_{\mathrm{I},g},p_{\mathrm{Q},g},m_g)\in
\mathcal U_g^{\mathrm{CGIM}}$ denote the code-domain tuple defined in
Sec.~\ref{sec:cgim_design}. SCGIM appends the $\log_2K_{\mathrm{t},g}$-bit port label
$\mathbf b_{\mathrm{t},g}$, which selects $k_g\in\{1,\ldots,K_{\mathrm{t},g}\}$. Hence,
$v_g=(u_g,k_g)\in\mathcal V_g^{\mathrm{SCGIM}}$ with
$\mathcal V_g^{\mathrm{SCGIM}}=\mathcal U_g^{\mathrm{CGIM}}\times
\{1,\ldots,K_{\mathrm{t},g}\}$. Each group therefore contains $P^2MK_{\mathrm{t},g}$ candidates
and conveys $B_{g}^{\mathrm{SCGIM}}=B_{g}^{\mathrm{CGIM}}+\log_2K_{\mathrm{t},g}$ bits,
yielding $B^{\mathrm{SCGIM}}=GB_{g}^{\mathrm{SCGIM}}$ bits per block.

The code-domain waveform $\mathbf x_g(u_g)$ is given by
\eqref{eq:cgim_group_signal}; SCGIM radiates this waveform through physical
port $\pi_g(k_g)$. The transmitted waveform matrix is
\begin{equation}
\mathbf X(\mathbf u)=
\left[\mathbf x_1^{\operatorname{T}}(u_1),\ldots,
\mathbf x_G^{\operatorname{T}}(u_G)\right]^{\operatorname{T}}
\in\mathbb C^{G\times L},
\label{eq:scgim_group_signal}
\end{equation}
where $\mathbf u=(u_1,\ldots,u_G)$. For
$g\in\{1,\ldots,G\}$ and $k\in\{1,\ldots,K_{\mathrm{t},g}\}$, define the
port-selection vector $\mathbf e_{\pi_g(k)}\in\mathbb B^{K_{\mathrm{t}}\times1}$
and the corresponding effective channel by
\begin{subequations}\label{eq:scgim_port_activation}
\begin{align}
[\mathbf e_{\pi_g(k)}]_{\ell}
&=\begin{cases}
1, & \ell=\pi_g(k),\\
0, & \ell\ne\pi_g(k),
\end{cases}
\quad \ell\in\{1,\ldots,K_{\mathrm{t}}\},
\label{eq:scgim_port_basis}\\
\mathbf h_{g,k}&=\mathbf H_{\mathrm{sel}}\mathbf e_{\pi_g(k)}
\in\mathbb C^{N_{\mathrm r}\times1}.
\label{eq:scgim_effective_channel}
\end{align}
\end{subequations}
The active-channel matrix is
$\mathbf H_{\mathrm{act}}(\mathbf k)=[\mathbf h_{1,k_1},\ldots,
\mathbf h_{G,k_G}]\in\mathbb C^{N_{\mathrm r}\times G}$, where
$\mathbf k=(k_1,\ldots,k_G)$ and
$\mathbf v=(v_1,\ldots,v_G)$. The received signal can then be written as
\begin{subequations}\label{eq:scgim_received_signal}
\begin{align}
\mathbf Y&=\mathbf H_{\mathrm{act}}(\mathbf k)
\mathbf X(\mathbf u)+\mathbf N
\label{eq:scgim_received_signal_matrix}\\
&=\sum_{g=1}^{G}\mathbf h_{g,k_g}\mathbf x_g(u_g)+\mathbf N
\in\mathbb C^{N_{\mathrm r}\times L}.
\label{eq:scgim_received_signal_sum}
\end{align}
\end{subequations}
\subsection{SCGIM Receiver Design}
The receiver applies the same group-wise despreading operation
$\mathbf R_g=\mathbf Y\mathbf C_g^{\operatorname{T}}$ as in
\eqref{eq:cgim_groupwise_despreading}. By code orthogonality, the despread
observation associated with code $p_{g,p}$ obeys
\eqref{eq:cgim_despread_output}, with $\mathbf h_{g,k_g}$ as the effective
channel. The inter-group components vanish under exact code orthogonality,
while the desired observation retains the coupled port index, code indices,
and QAM symbol.

\emph{ML detection:}
For $v=(u,k)\in\mathcal V_g^{\mathrm{SCGIM}}$, let
$\overline{\mathbf R}_g(v)=\mathbf h_{g,k}
(a_{\mathrm{I},m}\mathbf e_{p_{\mathrm{I}}}^{\operatorname T}
+\mathrm j a_{\mathrm{Q},m}\mathbf e_{p_{\mathrm{Q}}}^{\operatorname T})/\sqrt G$
denote the corresponding noiseless despread-domain response. The group-wise
ML detector jointly estimates all four
indices according to
\begin{equation}
\widehat v
=\left\langle
\widehat p_{\mathrm{I},g},\widehat p_{\mathrm{Q},g},
\widehat m_g,\widehat k_g
\right\rangle
=\argmin_{v\in\mathcal V_g^{\mathrm{SCGIM}}}
\left\|\mathbf R_g-\overline{\mathbf R}_g(v)\right\|_{\operatorname{F}}^2.
\label{eq:scgim_ml_detector}
\end{equation}
Accordingly, ML evaluates $P^2MK_{\mathrm{t},g}$ candidates in each group.

\emph{Greedy detection:}
GD first scores every candidate transmit port using all despread code
observations,
\begin{equation}
\Gamma_g(k)=\frac{\sum_{p=1}^{P}
|\mathbf h_{g,k}^{\operatorname{H}}\mathbf r_{g,p}|^2}{\|\mathbf h_{g,k}\|^2},
\qquad \widehat k_g=\argmax_k\Gamma_g(k).
\label{eq:scgim_gd_port_detection}
\end{equation}
Given $\widehat k_g$, the I/Q code indices and PAM amplitudes are
obtained from \eqref{eq:cgim_gd_index_detection} and
\eqref{eq:cgim_gd_qam_decision}, respectively, evaluated with the effective
channel $\mathbf h_{g,\widehat k_g}$.
\subsection{Cross-Domain Index Message-Passing Detection}
Sequential detection of the Tx-port and code-domain indices may discard their
cross-domain coupling. To retain this information, we develop a cross-domain
index message-passing detection (CD-IMPD) algorithm that propagates soft
information between the spatial-index and code-domain variable nodes over a
cycle-free factor graph. Hence, a single inward and outward message-passing sweep suffices to obtain all variable-node marginal beliefs.

\begin{figure}[!t]
    \centering
    \includegraphics[width=0.7\columnwidth]{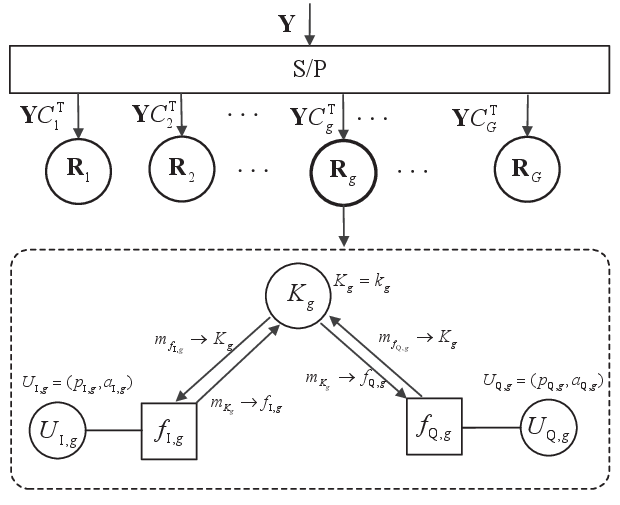}
    \vspace{-0.06in}
    \caption{Group-wise despreading and Tanner graph of the proposed CD-IMPD
    detector.}
    \label{fig:scgim_tanner_graph}
    \vspace{-0.2in}
\end{figure}

\emph{1) Factor-graph construction:}
As illustrated in Fig.~\ref{fig:scgim_tanner_graph}, for the $g$-th group, define three variable nodes
$U_{\mathrm{I},g}=(p_{\mathrm{I},g},a_{\mathrm{I},g})$,
$K_g=k_g$, and
$U_{\mathrm{Q},g}=(p_{\mathrm{Q},g},a_{\mathrm{Q},g})$.
$\mathcal D_\chi=\{1,\ldots,P\}\times\mathcal A$ denotes the state space
of branch $\chi\in\{\mathrm{I},\mathrm{Q}\}$. Let
$\Lambda_{\chi,g}(k,p,a)$ denote
the local likelihood associated with branch $\chi$, candidate port $k$, and
branch state $(p,a)$. Conditioned on $K_g$, the I/Q despread observations
are independent; hence, the joint posterior factorizes as
\begin{equation}
p(k,p_{\mathrm{I}},a_{\mathrm{I}},p_{\mathrm{Q}},a_{\mathrm{Q}}
\mid\mathbf R_g)\propto
\Lambda_{\mathrm{I},g}(k,p_{\mathrm{I}},a_{\mathrm{I}})
\Lambda_{\mathrm{Q},g}(k,p_{\mathrm{Q}},a_{\mathrm{Q}}).
\label{eq:scgim_spa_factorization}
\end{equation}
\vspace{-0.2in}

Accordingly, the factor graph is
$U_{\mathrm{I},g}$--$f_{\mathrm{I},g}$--$K_g$--$f_{\mathrm{Q},g}$--$U_{\mathrm{Q},g}$,
where $f_{\mathrm{I},g}$ and $f_{\mathrm{Q},g}$ are the I/Q observation factors.
The inward and outward information flows are
$U_{\mathrm{I},g}\rightarrow f_{\mathrm{I},g}\rightarrow K_g
\leftarrow f_{\mathrm{Q},g}\leftarrow U_{\mathrm{Q},g}$ and
$U_{\mathrm{I},g}\leftarrow f_{\mathrm{I},g}\leftarrow K_g
\rightarrow f_{\mathrm{Q},g}\rightarrow U_{\mathrm{Q},g}$, respectively.
For $(p,a)\in\mathcal D_\chi$, define the candidate-dependent matched statistic
$z_{g,k,p}=\mathbf h_{g,k}^{\operatorname{H}}\mathbf r_{g,p}$. The
two branch distances are
\begin{subequations}\label{eq:scgim_spa_branch_metric}
\begin{align}
d_{\mathrm{I},g}(k,p,a)
&=\frac{\|\mathbf h_{g,k}\|^2}{G}a^2
-\frac{2a}{\sqrt G}\Re\{z_{g,k,p}\},
\label{eq:scgim_spa_branch_metric_i}\\
d_{\mathrm{Q},g}(k,p,a)
&=\frac{\|\mathbf h_{g,k}\|^2}{G}a^2
-\frac{2a}{\sqrt G}\Im\{z_{g,k,p}\}.
\label{eq:scgim_spa_branch_metric_q}
\end{align}
\end{subequations}
The corresponding branch factors are $\Lambda_{\chi,g}(k,p,a)=
\exp[-d_{\chi,g}(k,p,a)/N_0]$.

\emph{2) Cross-domain message updates:}
Under uniform signaling, the constant variable-to-factor messages can be
absorbed into normalization. The inward message to the spatial port node is
\begin{equation}
\mu_{f_{\chi,g}\rightarrow K_g}(k)
\propto\sum_{(p,a)\in\mathcal D_\chi}
\Lambda_{\chi,g}(k,p,a).
\label{eq:scgim_spa_forward_message}
\end{equation}
The port node fuses the two incoming messages according to
\begin{equation}
b_{K_g}(k)\propto
\mu_{f_{\mathrm{I},g}\rightarrow K_g}(k)
\mu_{f_{\mathrm{Q},g}\rightarrow K_g}(k).
\label{eq:scgim_spa_port_belief}
\end{equation}
Rather than slicing $b_{K_g}(k)$, the outward pass returns to each branch the
extrinsic message received from the opposite branch,
\begin{subequations}\label{eq:scgim_spa_port_outgoing}
\begin{align}
\mu_{K_g\rightarrow f_{\mathrm{I},g}}(k)&=\mu_{f_{\mathrm{Q},g}\rightarrow K_g}(k),
\label{eq:scgim_spa_port_outgoing_I}\\
\mu_{K_g\rightarrow f_{\mathrm{Q},g}}(k)&=\mu_{f_{\mathrm{I},g}\rightarrow K_g}(k).
\label{eq:scgim_spa_port_outgoing_Q}
\end{align}
\end{subequations}
Combining this message with the local likelihood and marginalizing the
candidate ports gives
\begin{equation}
\mu_{f_{\chi,g}\rightarrow U_{\chi,g}}(p,a)
=\sum_{k=1}^{K_{\mathrm{t},g}}\Lambda_{\chi,g}(k,p,a)
\mu_{K_g\rightarrow f_{\chi,g}}(k).
\label{eq:scgim_spa_backward_message}
\end{equation}

\emph{3) Marginal decisions:}
The beliefs and marginal maximum a posteriori (MAP) decisions of all three variable nodes are
\begin{subequations}\label{eq:scgim_spa_branch_decision}
\begin{align}
\widehat k_g&=\argmax_k b_{K_g}(k),
\label{eq:scgim_spa_port_decision}\\
b_{U_{\chi,g}}(p,a)&\propto
\mu_{f_{\chi,g}\rightarrow U_{\chi,g}}(p,a),
\label{eq:scgim_spa_branch_belief}\\
\left\langle\widehat p_{\chi,g},\widehat a_{\chi,g}\right\rangle
&=\argmax_{(p,a)\in\mathcal D_\chi}b_{U_{\chi,g}}(p,a).
\label{eq:scgim_spa_branch_map}
\end{align}
\end{subequations}
Finally, the I/Q branch decisions provide the two code indices and PAM
amplitudes, from which $\widehat m_g$ is reconstructed.

\section{Spectral Efficiency and Detection Complexity Analysis}
\label{sec:se_complexity}

\subsection{Spectral Efficiency Analysis}
Table~\ref{tab:spectral_efficiency_comparison} compares the information-bearing
domains and spectral efficiencies of the proposed schemes with representative
code- and spatial-domain IM counterparts. Here, SIM denotes spatial-domain IM;
SM, QSM, FAIM, and FAG-IM denote spatial modulation, quadrature spatial
modulation, fluid antenna index modulation, and grouped FA-IM, respectively,
whereas GCIM-SM and CIM-QSM denote GCIM combined with SM and CIM-aided QSM.
$N_{\mathrm t}$ is the number
of simultaneously active transmit ports/antennas in multiport FA-IM. The Rx-FAS performs receive-port selection and therefore does not convey any additional index bits.

\begin{table}[!t]
\caption{Design Features and Spectral Efficiencies of Representative IM Schemes}
\label{tab:spectral_efficiency_comparison}
\centering
\scriptsize
\renewcommand{\arraystretch}{1.08}
\setlength{\tabcolsep}{1.5pt}
\begin{tabular}{@{}lcccc@{\hspace{3pt}}l@{}}
\toprule
Scheme & \shortstack{Tx-\\FA} & \shortstack{Rx-\\FA} & CIM & SIM
& Bits per codeword symbol \\
\midrule
CIM~\cite{Kaddoum2015CIM}
& -- & -- & $\checkmark$ & -- & $\log_2P+\log_2M$ \\
GCIM~\cite{Kaddoum2016GCIM}
& -- & -- & $\checkmark$ & -- & $2\log_2P+\log_2M$ \\
QCIM~\cite{Cai2025QCIM}
& -- & -- & $\checkmark$ & -- & $4\log_2P+2\log_2M$ \\
\addlinespace
Proposed CGIM
& -- & $\checkmark$ & $\checkmark$ & --
& $G\left(2\log_2P+\log_2M\right)$ \\
\addlinespace
SM~\cite{Mesleh2008SM}
& -- & -- & -- & $\checkmark$ & $\log_2N_{\mathrm t}+\log_2M$ \\
QSM~\cite{Mesleh2015QSM}
& -- & -- & -- & $\checkmark$ & $2\log_2N_{\mathrm t}+\log_2M$ \\
\addlinespace
FAIM~\cite{Chen2024FAIM,Faddoul2025CodedFAIM}
& $\checkmark$ & -- & -- & $\checkmark$
& $\log_2K_{\mathrm t}+\log_2M$ \\
Multiport FA-IM~\cite{Zhu2024FAIM}
& $\checkmark$ & -- & -- & $\checkmark$
& $\left\lfloor\log_2\binom{K_{\mathrm t}}{N_{\mathrm t}}\right\rfloor
  +N_{\mathrm t}\log_2M$ \\
FAG-IM~\cite{Guo2025FluidAntennaIndexModulation}
& $\checkmark$ & -- & -- & $\checkmark$
& $G\left(\log_2K_{\mathrm t,g}+\log_2M\right)$ \\
\addlinespace
GCIM-SM~\cite{Cogen2021GCIMSM}
& -- & -- & $\checkmark$ & $\checkmark$
& $\log_2N_{\mathrm t}+2\log_2P+\log_2M$ \\
CIM-QSM~\cite{Aydin2019CIMQSM}
& -- & -- & $\checkmark$ & $\checkmark$
& $2\log_2N_{\mathrm t}+2\log_2P+\log_2M$ \\
\addlinespace
Proposed SCGIM
& $\checkmark$ & $\checkmark$ & $\checkmark$ & $\checkmark$
& $G\left(\log_2K_{\mathrm t,g}+2\log_2P+\log_2M\right)$ \\
\bottomrule
\end{tabular}
\end{table}

\subsection{Detection Complexity Analysis}
\label{subsec:detection_complexity}
Complexity is measured by the number of real multiplications (RMs) per
codeword block. A complex multiplication and a squared magnitude incur four
and two RMs, respectively. Codebook- and constellation-dependent quantities
are precomputed; additions, comparisons, and index operations are omitted.

The group-wise despreading operation in \eqref{eq:cgim_groupwise_despreading}
computes $GP$ length-$L$
real-code correlations for each of the $N_{\mathrm r}$ observations and hence
incurs $C_{\mathrm{desp}}^{\mathrm{RM}}=2GN_{\mathrm r}PL$ real multiplications (RMs). Since this
operation is common to all detectors, it is excluded from
Table~\ref{tab:detection_complexity}. Under direct implementation, the squared
norm of an $n$-dimensional complex vector costs $2n$ RMs, whereas the product
of an $m\times n$ complex matrix and an $n\times p$ complex matrix costs
$4mnp$ RMs. Consequently, $\|\mathbf h\|^2$ and
$\mathbf h^H\mathbf R_g$ require $2N_{\mathrm r}$ and $4N_{\mathrm r}P$ RMs
per candidate channel, respectively. These candidate-independent quantities
are reused in all subsequent metric evaluations. The resulting
detector-specific counts are given in Table~\ref{tab:detection_complexity}.

\begin{table}[t]
\caption{Real-Multiplication Complexities of the Considered Detectors}
\label{tab:detection_complexity}
\centering
\scriptsize
\renewcommand{\arraystretch}{1.12}
\setlength{\tabcolsep}{3.2pt}
\begin{tabular}{@{}ll@{}}
\toprule
Detector & RMs per codeword block \\
\midrule
CGIM-ML
& $G\left(2N_{\mathrm r}+4N_{\mathrm r}P+3P^2M\right)$ \\
CGIM-GD
& $G\left(2N_{\mathrm r}+4N_{\mathrm r}P+2\right)$ \\
SCGIM-ML
& $GK_{\mathrm{t},g}\left(2N_{\mathrm r}+4N_{\mathrm r}P+3P^2M\right)$ \\
SCGIM-GD
& $G\left[K_{\mathrm{t},g}\left(2N_{\mathrm r}+4N_{\mathrm r}P+2P\right)+2\right]$ \\
SCGIM-CD-IMPD
& $GK_{\mathrm{t},g}\left(2N_{\mathrm r}+4N_{\mathrm r}P+6P\sqrt M\right)$ \\
\bottomrule
\end{tabular}
\end{table}

The exact I/Q decomposition eliminates the $P^2M$ joint enumeration in
CGIM-GD. For SCGIM, exhaustive ML detection introduces an additional factor
$K_{\mathrm{t},g}$ due to the Tx-FA port search. SCGIM-GD suppresses this joint
search through a hard port decision, whereas CD-IMPD preserves spatial--code
coupling while replacing each $P^2M$ search by two $P\sqrt M$ branchwise state
spaces. Exponential and logarithmic evaluations used for CD-IMPD message
normalization are excluded from the RM count.
\vspace{-0.1in}
\section{BER Performance Analysis}
\label{sec:theoretical_analysis}

This section analyzes the BER performance of the proposed CGIM and SCGIM
schemes. We first derive the CGIM BER from its code-domain decision regions and
then develop a BER analysis based on pairwise error probability (PEP) for SCGIM.

\subsection{CGIM Performance Analysis}
The CGIM analysis proceeds from the code-domain ML error event to the
unconditional BER. We first derive the exact conditional BER for a fixed
combining gain $\Omega=\omega$, characterize the PDF $f_\Omega(\omega)$ of the
Rx-FAS selected gain, and then average over this gain to obtain
$P_{\mathrm b}^{\mathrm{CGIM}}(\gamma_{\mathrm b})$. The transmitted state of group $g$ is
$u_g=\langle p_{\mathrm I,g},p_{\mathrm Q,g},m_g\rangle
\in\mathcal U_g^{\mathrm{CGIM}}$, and its competing event is
$\mathcal E_g(u_g\!\rightarrow\!\widehat u_g)
=\{D_g(\widehat u_g)<D_g(u_g)\}$. Since the codebooks assigned to different
groups are mutually orthogonal, the overall ML metric decomposes into $G$
group metrics. Moreover, the I/Q branches have identical conditional error
statistics. It is therefore sufficient to analyze one real-valued branch of
an arbitrary group.

\emph{1) Conditional error probability:}
Let the ordered normalized PAM alphabet be
$\overline{\mathcal A}=\{\alpha_i\}_{i=1}^{\sqrt M}$, where
$\alpha_i=a_i/\sqrt G$. Conditioned on the effective combining gain
$\Omega=\|\mathbf h_{\mathrm{sel}}\|^2=\omega$, the despread observation of
the transmitted code follows $\mathcal N(\alpha_i,\sigma^2(\omega))$, whereas
each unselected-code observation follows $\mathcal N(0,\sigma^2(\omega))$,
with $\sigma^2(\omega)=N_0/(2\omega)$. For a candidate amplitude $\alpha_j$,
its ML score and decision region are
\begin{subequations}\label{eq:cgim_theory_score_region}
\begin{align}
\vartheta_j(x)&=2\alpha_jx-\alpha_j^2,
\label{eq:cgim_theory_score}\\
\mathcal R_j&=[\tau_{j-1},\tau_j),\qquad
\tau_j=\frac{\alpha_j+\alpha_{j+1}}{2},
\label{eq:cgim_theory_region}
\end{align}
\end{subequations}
where $\tau_0=-\infty$ and $\tau_{\sqrt M}=+\infty$.

For $X\sim\mathcal N(\bar\alpha,\sigma^2)$ and a real threshold $\kappa$, define
the CDF of the maximum PAM score as
\begin{equation}
F_{\vartheta}(\kappa;\bar\alpha,\sigma)
=\Pr\!\left\{\max_{\alpha\in\overline{\mathcal A}}
(2\alpha X-\alpha^2)\leq\kappa\right\}.
\label{eq:cgim_score_cdf_definition}
\end{equation}
Let
$\ell(\kappa)=\max_{\alpha<0}(\kappa+\alpha^2)/(2\alpha)$ and
$u(\kappa)=\min_{\alpha>0}(\kappa+\alpha^2)/(2\alpha)$. Then
\begin{equation}
F_{\vartheta}(\kappa;\bar\alpha,\sigma)
=\left[\Phi\!\left(\frac{u(\kappa)-\bar\alpha}{\sigma}\right)
-\Phi\!\left(\frac{\ell(\kappa)-\bar\alpha}{\sigma}\right)\right]^+,
\label{eq:cgim_score_cdf_closed}
\end{equation}
where $[x]^+=\max\{x,0\}$ and $\Phi(\cdot)$ is the standard normal CDF.
Let $\phi_\sigma(x-\bar\alpha)$ denote the
$\mathcal N(\bar\alpha,\sigma^2)$ PDF.

\begin{figure}[!t]
    \centering
    \includegraphics[width=\columnwidth]{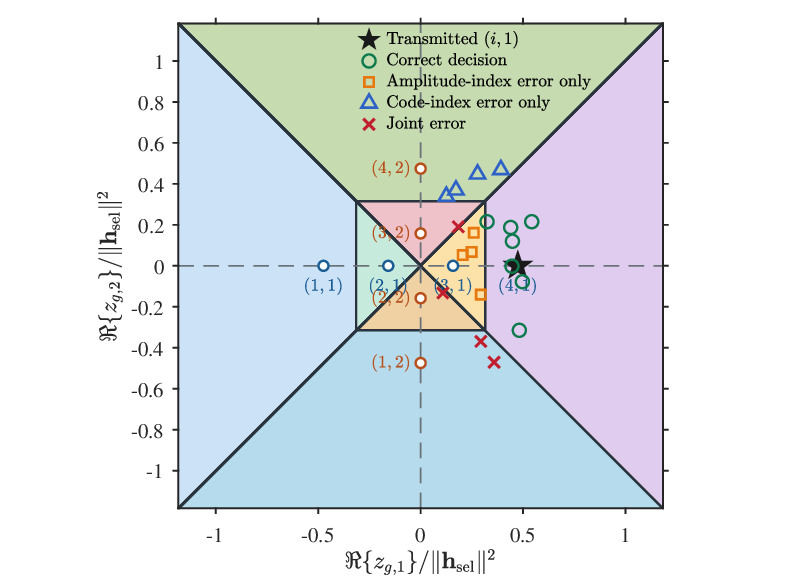}
    \vspace{-0.20in}
    \caption{Two-dimensional code-domain ML decision regions for $P=2$ and
    $M=16$.}
    \label{fig:cgim_code_domain_regions}
    \vspace{-0.20in}
\end{figure}

Renumber the $P$ codes within a group as $p\in\{1,\ldots,P\}$. By code-index
symmetry, assume that code 1 is transmitted with amplitude $\alpha_i$. The
probabilities of deciding $(j,1)$ and $(j,p)$, $p\ne1$, are, respectively,
\begin{subequations}\label{eq:cgim_conditional_decision_events}
\begin{align}
\Pi_{i,j}^{(1)}(\sigma)
&=\int_{\mathcal R_j}\!\phi_\sigma(x-\alpha_i)
F_{\vartheta}(\vartheta_j(x);0,\sigma)^{P-1}\,\mathrm dx,
\label{eq:cgim_conditional_correct_code}\\
\Pi_{i,j}^{(p)}(\sigma)
&=\int_{\mathcal R_j}\!\phi_\sigma(x)
F_{\vartheta}(\vartheta_j(x);0,\sigma)^{P-2}
F_{\vartheta}(\vartheta_j(x);\alpha_i,\sigma)\,\mathrm dx.
\label{eq:cgim_conditional_wrong_code}
\end{align}
\end{subequations}
The two events form a mutually exclusive partition of the ML decision region:
\eqref{eq:cgim_conditional_correct_code} retains the transmitted code, whereas
\eqref{eq:cgim_conditional_wrong_code} selects an erroneous code. Thus,
\eqref{eq:cgim_conditional_decision_events} accounts for the simultaneous
competition among all $P$ despread code outputs without a pairwise-error
approximation. Fig.~\ref{fig:cgim_code_domain_regions} illustrates this
geometry for $P=2$ and $M=16$, where each colored region represents a joint
code-index and PAM-amplitude decision event. For $P>2$, the normalized
despread outputs form an observation vector in $\mathbb R^P$, and the region
associated with decision $(j,p)$ is the intersection of the affine half-spaces
$\vartheta_j(x_p)\geq\vartheta_{j'}(x_{p'})$ for all competing
$(j',p')$, hence forming a convex polyhedron. Its Gaussian probability is
evaluated by \eqref{eq:cgim_conditional_decision_events} through the
maximum-score CDF, avoiding direct integration over the $P$-dimensional
polyhedron.

Let $\mathbf b_{i,p}$ be the branch bit label of amplitude $i$ and code index
$p$. With $1\leq i,j\leq\sqrt M$ and $1\leq p\leq P$, the exact conditional
BER kernel is
\begin{equation}
\mathcal K_{\mathrm{CGIM}}(\omega,\gamma_{\mathrm b})
\!=\!\frac{\sum_{i,j,p}d_{\operatorname H}(\mathbf b_{i,1},\mathbf b_{j,p})
\Pi_{i,j}^{(p)}\!\left(\sqrt{\frac{N_0}{2\omega}}\right)}
{\sqrt M(\log_2P+\frac12\log_2M)},
\label{eq:cgim_conditional_ber}
\end{equation}
where $d_{\operatorname H}(\cdot,\cdot)$ denotes Hamming distance and
$N_0=E_{\mathrm s}/(B^{\mathrm{CGIM}}\gamma_{\mathrm b})$. By branch and group
symmetry, \eqref{eq:cgim_conditional_ber} also gives the conditional BER of the
complete CGIM block.

\emph{2) Rx-FAS selection statistics:}
For receive RF chain $r$, define the selected-port power gain $\Omega_r$ and total combining
gain $\Omega$ as
\begin{subequations}\label{eq:cgim_selected_gains}
\begin{align}
\Omega_r&=\max_{1\le q\le K_r}|h_{r,q}|^2,
\label{eq:cgim_single_rxfa_gain}\\
\Omega&=\sum_{r=1}^{N_{\mathrm r}}\Omega_r=\|\mathbf h_{\mathrm{sel}}\|^2.
\label{eq:cgim_total_rxfa_gain}
\end{align}
\end{subequations}
The CDF
$F_{\Omega_r}(\omega)=
\Pr\{\max_{1\leq q\leq K_r}|h_{r,q}|^2\leq\omega\}
=\Pr\{\bigcap_{q=1}^{K_r}|h_{r,q}|^2\leq\omega\}$
is the probability that every port in the $r$-th receive-port set remains
below the gain threshold $\omega$.

\emph{a) Correlated Rayleigh fading:}
Let $Q_\nu(a,b)$ denote the generalized Marcum-$Q$ function of order $\nu$
and $\overline Q_\nu(a,b)=1-Q_\nu(a,b)$ its complement.
Under the reference-port model in \eqref{eq:rx_reference_model}, conditioning
on the reference-port power $|h_{r,1}|^2=t$ gives
\cite{Wong2021FluidAntennaSystems}
\begin{equation}
\begin{aligned}
F_{\Omega_r}^{\mathrm{Ray}}(\omega)
&=\int_0^\omega e^{-t}\prod_{q=2}^{K_r}
\overline Q_1\!\left(
\sqrt{\frac{2\mu_q^2t}{1-\mu_q^2}},
\sqrt{\frac{2\omega}{1-\mu_q^2}}
\right)\mathrm dt.
\end{aligned}
\label{eq:cgim_rayleigh_selected_gain_cdf}
\end{equation}

\emph{b) Correlated Nakagami-$m$ fading:}
For a positive integer $m$, the Gaussian-component construction above yields
unit-power Nakagami-$m$ fading
\cite{VegaSanchez2024SimpleMethod,Tlebaldiyeva2022EnhancingQoS}.
Here, $\operatorname{Gamma}(\alpha,\theta)$ denotes the Gamma distribution
with shape $\alpha$ and scale $\theta$, whereas $\Gamma(\cdot)$ is the Gamma
function. Since
$T=m|h_{r,1}|^2\sim\operatorname{Gamma}(m,1)$, reference-port conditioning
yields
\begin{equation}
\begin{aligned}
F_{\Omega_r}^{\mathrm{Nak}}(\omega)
&=\frac{1}{\Gamma(m)}\int_0^{m\omega}t^{m-1}e^{-t}
\\[-0.4ex]
&\quad{}\times\prod_{q=2}^{K_r}\overline Q_m\!\left(
\sqrt{\frac{2\mu_q^2t}{1-\mu_q^2}},
\sqrt{\frac{2m\omega}{1-\mu_q^2}}
\right)\mathrm dt.
\end{aligned}
\label{eq:cgim_nakagami_selected_gain_cdf}
\end{equation}
Setting $m=1$ recovers \eqref{eq:cgim_rayleigh_selected_gain_cdf}.

\emph{3) Average BER:}
Let $f_{\Omega_r}(\omega)=\mathrm dF_{\Omega_r}(\omega)/\mathrm d\omega$.
Since the $N_{\mathrm r}$ receive-port sets are mutually independent, the PDF of the total
gain in \eqref{eq:cgim_total_rxfa_gain} is
\begin{equation}
f_\Omega=f_{\Omega_1}*f_{\Omega_2}*\cdots*f_{\Omega_{N_{\mathrm r}}}.
\label{eq:cgim_total_gain_pdf}
\end{equation}
Here, the ordinary convolution is defined as
$(f_X*f_Y)(\omega)=\int_0^\omega f_X(t)f_Y(\omega-t)\,\mathrm dt$.
Combining the conditional BER kernel in \eqref{eq:cgim_conditional_ber} with
the selected-gain PDF gives the desired CGIM BER,
\begin{equation}
P_{\mathrm b}^{\mathrm{CGIM}}(\gamma_{\mathrm b})
=\int_0^\infty\mathcal K_{\mathrm{CGIM}}(\omega,\gamma_{\mathrm b})
f_\Omega(\omega)\,\mathrm d\omega.
\label{eq:cgim_average_ber}
\end{equation}
This result separates the code-domain decision geometry, contained in
$\mathcal K_{\mathrm{CGIM}}$, from the channel dependence, represented by
$f_\Omega$, and requires no channel-realization sampling. For
fixed-position-antenna reception (FPA-Rx),
$K_r=1$; under Rayleigh fading,
\begin{equation}
F_\Omega^{\mathrm{FPA}}(\omega)
=1-e^{-\omega}\sum_{n=0}^{N_{\mathrm r}-1}\frac{\omega^n}{n!}.
\label{eq:cgim_fpa_total_gain_cdf}
\end{equation}

\emph{4) AWGN special case:}
For the normalized AWGN channel, $|h_{r,q}|^2=1$, and hence
$\Omega_r=1$ and $\Omega=N_{\mathrm r}$. Accordingly,
\begin{equation}
P_{\mathrm b}^{\mathrm{CGIM,AWGN}}(\gamma_{\mathrm b})
=\mathcal K_{\mathrm{CGIM}}(N_{\mathrm r},\gamma_{\mathrm b}).
\label{eq:cgim_awgn_ber}
\end{equation}
Thus, FA-Rx and FPA-Rx have identical performance in AWGN under the same
RF-chain count and power normalization.

\emph{Remark 1 (Rayleigh-Series Representation):}
Alternatively, the Rayleigh-fading result admits an infinite-series
representation. Define the regularized lower incomplete Gamma function as
$P_\gamma(s,x)=\gamma(s,x)/\Gamma(s)$, where $\gamma(\cdot,\cdot)$ is the
lower incomplete Gamma function. The complementary Marcum-$Q$ factor in
\eqref{eq:cgim_rayleigh_selected_gain_cdf} admits
\begin{equation}
\begin{aligned}
&\overline Q_1\!\left(
\sqrt{\frac{2\mu_q^2t}{1-\mu_q^2}},
\sqrt{\frac{2\omega}{1-\mu_q^2}}
\right)\\[-0.4ex]
&\quad=e^{-\mu_q^2t/(1-\mu_q^2)}\sum_{n=0}^{\infty}
\frac{[\mu_q^2t/(1-\mu_q^2)]^n}{n!}
P_\gamma\!\left(n+1,\frac{\omega}{1-\mu_q^2}\right).
\end{aligned}
\label{eq:cgim_marcum_series}
\end{equation}
Let $\mathbb N_0=\{0,1,\ldots\}$ and
$\mathbf n=(n_2,\ldots,n_{K_r})\in\mathbb N_0^{K_r-1}$,
$N_{\mathbf n}=\sum_{q=2}^{K_r}n_q$, and
$\delta=1+\sum_{q=2}^{K_r}\mu_q^2/(1-\mu_q^2)$. Applying
\eqref{eq:cgim_marcum_series} to
\eqref{eq:cgim_rayleigh_selected_gain_cdf} yields
\begin{subequations}\label{eq:cgim_rxfa_gain_series}
\begin{align}
F_{\Omega_r}(\omega)&=\sum_{\mathbf n\in\mathbb N_0^{K_r-1}}
\mathcal T_{\mathbf n}(\omega),
\label{eq:cgim_rxfa_gain_series_sum}\\
\mathcal T_{\mathbf n}(\omega)
&=\frac{\gamma(N_{\mathbf n}+1,\delta\omega)}{\delta^{N_{\mathbf n}+1}}
\notag\\[-0.4ex]
&\quad{}\times
\prod_{q=2}^{K_r}\frac{[\mu_q^2/(1-\mu_q^2)]^{n_q}}{n_q!}
P_\gamma\!\left(n_q+1,\frac{\omega}{1-\mu_q^2}\right).
\label{eq:cgim_rxfa_gain_series_term}
\end{align}
\end{subequations}
The infinite series is equivalent to the Marcum-$Q$ integral. Its finite-order
truncation converges to \eqref{eq:cgim_rayleigh_selected_gain_cdf} as the
retained order increases. Hence, either representation can be substituted into
\eqref{eq:cgim_total_gain_pdf} and \eqref{eq:cgim_average_ber} to evaluate
$P_{\mathrm b}^{\mathrm{CGIM}}(\gamma_{\mathrm b})$.

\subsection{SCGIM Performance Analysis}
Unlike CGIM, the ML error region of the proposed SCGIM is coupled by the
correlation among Tx-FA ports and thus cannot be decomposed solely through
code orthogonality. We therefore adopt the classical PEP framework for
index-modulation analysis
\cite{Guo2025FluidAntennaIndexModulation}.

\emph{1) Conditional PEP:}
For $v=(u,k)\in\mathcal V_g^{\mathrm{SCGIM}}$, define its
physical-port-domain signal and the difference between an ordered candidate
pair as
\begin{subequations}\label{eq:scgim_pair_signals}
\begin{align}
\widetilde{\mathbf X}_g(v)
&=\mathbf e_{\pi_g(k)}\mathbf x_g(u)
\in\mathbb C^{K_{\mathrm t}\times L},
\label{eq:scgim_physical_signal}\\
\boldsymbol{\Delta}_{v,\widehat v}
&=\widetilde{\mathbf X}_g(v)-\widetilde{\mathbf X}_g(\widehat v).
\label{eq:scgim_pair_difference}
\end{align}
\end{subequations}
Define the selected squared Euclidean distance of the ordered pair as
$\Gamma_{v,\widehat v}=\|\mathbf H_{\mathrm{sel}}
\boldsymbol{\Delta}_{v,\widehat v}\|_{\operatorname F}^{2}
=\sum_{r=1}^{N_{\mathrm r}}Z_{r,v,\widehat v}$. Conditioned on
$\mathbf H_{\mathrm{sel}}$, the PEP of $v\rightarrow\widehat v$ is
\begin{equation}
P(v\rightarrow\widehat v\mid\mathbf H_{\mathrm{sel}})
=Q\!\left(\sqrt{\frac{\Gamma_{v,\widehat v}}{2N_0}}\right).
\label{eq:scgim_conditional_pep}
\end{equation}

\emph{2) Selected quadratic-form transform:}
Let $\mathbf z_{r,q}\sim\mathcal{CN}(\mathbf 0,\mathbf I_{K_{\mathrm t}})$
denote the independent Gaussian vector underlying the channel response of the $q$-th
candidate receive port, such that $\mathbf h_{r,q}=\mathbf z_{r,q}\mathbf
R_{\mathrm t}^{1/2}$, and let
$\mathbf z_r^\star=\mathbf z_{r,q_r^\star}$. For an ordered pair
$(v,\widehat v)$, define
\begin{subequations}\label{eq:scgim_quadratic_form}
\begin{align}
\mathbf B_{v,\widehat v}
&=\mathbf R_{\mathrm t}^{1/2}\mathbf A_{v,\widehat v}
\mathbf R_{\mathrm t}^{1/2},
\label{eq:scgim_error_matrix}\\
Z_{r,v,\widehat v}
&=\mathbf z_r^\star\mathbf B_{v,\widehat v}
(\mathbf z_r^\star)^{\operatorname H},\qquad
G_r^\star=\mathbf z_r^\star\mathbf R_{\mathrm t}
(\mathbf z_r^\star)^{\operatorname H}.
\label{eq:scgim_distance_and_gain}
\end{align}
\end{subequations}
Thus, the pairwise distance and the Rx-FAS selection statistic are retained
as two quadratic forms of the same Gaussian vector instead of being separated
into independent gain and direction variables.

Before port selection, $G_{r,q}=\|\mathbf h_{r,q}\|^2$ is represented by
the moment-matched Gamma distribution
$\operatorname{Gamma}(m_{\mathrm t},\theta_{\mathrm t})$
\cite{Satterthwaite1946}, where
\begin{equation}
m_{\mathrm t}=\frac{\operatorname{tr}^{2}(\mathbf R_{\mathrm t})}
{\operatorname{tr}(\mathbf R_{\mathrm t}^{2})},\qquad
\theta_{\mathrm t}=\frac{\operatorname{tr}(\mathbf R_{\mathrm t}^{2})}
{\operatorname{tr}(\mathbf R_{\mathrm t})}.
\label{eq:scgim_tx_energy_gamma}
\end{equation}
Under the reference-correlation model, the selected channel-energy CDF is
\begin{equation}
\begin{aligned}
F_{G_r^\star}(g)
&=\int_0^g\frac{x^{m_{\mathrm t}-1}e^{-x/\theta_{\mathrm t}}}
{\Gamma(m_{\mathrm t})\theta_{\mathrm t}^{m_{\mathrm t}}}
\\[-0.4ex]
&\quad\times\prod_{q=2}^{K_r}\overline Q_{m_{\mathrm t}}\!\left(
\sqrt{\frac{2\mu_q^2x}{(1-\mu_q^2)\theta_{\mathrm t}}},
\sqrt{\frac{2g}{(1-\mu_q^2)\theta_{\mathrm t}}}
\right)\mathrm dx.
\end{aligned}
\label{eq:scgim_selected_row_energy_cdf}
\end{equation}
Let $f_{G_r^\star}(g)=\mathrm dF_{G_r^\star}(g)/\mathrm dg$ and let
$f_G(g)$ denote the Gamma density specified by
\eqref{eq:scgim_tx_energy_gamma}. The selection-induced change of measure
is represented by the energy-domain density ratio
\begin{equation}
\omega(g)=\frac{f_{G_r^\star}(g)}{f_G(g)}.
\label{eq:scgim_selection_density_ratio}
\end{equation}
Accordingly, the one-branch transform is written as
$\mathcal L_{Z_{v,\widehat v}}(s)\simeq
\mathbb E[e^{-s\mathbf z\mathbf B_{v,\widehat v}\mathbf z^{\operatorname H}}
\omega(\mathbf z\mathbf R_{\mathrm t}\mathbf z^{\operatorname H})]$.

For $s\geq0$, define
$\mathbf C_{v,\widehat v}(s)=(\mathbf I+s\mathbf
B_{v,\widehat v})^{-1}$. Gaussian exponential tilting gives the base
quadratic-form transform $\det(\mathbf I+s\mathbf
B_{v,\widehat v})^{-1}$ and changes the covariance of the underlying Gaussian
vector
to $\mathbf C_{v,\widehat v}(s)$. Under this tilted measure, the first two
moments of $G_s=\mathbf z_s\mathbf R_{\mathrm t}
\mathbf z_s^{\operatorname H}$ are
\begin{subequations}\label{eq:scgim_tilted_gain_moments}
\begin{align}
\mathbb E[G_s]
&=\operatorname{tr}[\mathbf C_{v,\widehat v}(s)\mathbf R_{\mathrm t}],\\
\operatorname{Var}(G_s)
&=\operatorname{tr}\!\left\{
[\mathbf C_{v,\widehat v}(s)\mathbf R_{\mathrm t}]^2\right\}.
\end{align}
\end{subequations}
Matching these moments by
$G_s\sim\operatorname{Gamma}(m_s,\theta_s)$, where
$m_s=\mathbb E^2[G_s]/\operatorname{Var}(G_s)$ and
$\theta_s=\operatorname{Var}(G_s)/\mathbb E[G_s]$, yields
\begin{equation}
\mathcal L_{Z_{v,\widehat v}}(s)
\simeq\frac{1}{\det(\mathbf I+s\mathbf B_{v,\widehat v})}
\int_0^\infty\omega(g)
\frac{g^{m_s-1}e^{-g/\theta_s}}
{\Gamma(m_s)\theta_s^{m_s}}\,\mathrm dg.
\label{eq:scgim_pair_laplace}
\end{equation}
All expectations in \eqref{eq:scgim_pair_laplace} are evaluated by
deterministic quadrature; no channel realizations are generated. For
$K_r=1$, $f_{G_r^\star}=f_G$ and hence $\omega(g)=1$, so
\eqref{eq:scgim_pair_laplace} reduces to the exact FPA result
$\det(\mathbf I+s\mathbf B_{v,\widehat v})^{-1}$.

\emph{3) Average PEP and BER:}
Since the $N_{\mathrm r}$ receive-port sets are mutually independent,
$\mathcal L_{\Gamma_{v,\widehat v}}(s)=
[\mathcal L_{Z_{v,\widehat v}}(s)]^{N_{\mathrm r}}$. The exact UPEP is
$P_{\mathrm U}(v\!\rightarrow\!\widehat v)=
\mathbb E[Q(\sqrt{\Gamma_{v,\widehat v}/(2N_0)})]$. Substituting the
Laplace approximation in \eqref{eq:scgim_pair_laplace} into Craig's
representation gives
\begin{equation}
\widehat P_{\mathrm C}(v\rightarrow\widehat v)
\simeq\frac{1}{\pi}\int_0^{\pi/2}
\left[\mathcal L_{Z_{v,\widehat v}}\!\left(
\frac{1}{4N_0\sin^2\theta}\right)\right]^{N_{\mathrm r}}\mathrm d\theta.
\label{eq:scgim_average_pep_craig}
\end{equation}
For numerical evaluation, applying the classical approximation
$Q(x)\simeq \frac{1}{12}\exp(-x^2/2)+\frac{1}{4}\exp(-2x^2/3)$
\cite{Chiani2003QFunction} to the conditional PEP yields
\begin{equation}
\widetilde P(v\rightarrow\widehat v)
=\frac{1}{12}\left[\mathcal L_{Z_{v,\widehat v}}\!\left(
\frac{1}{4N_0}\right)\right]^{N_{\mathrm r}}
+\frac{1}{4}\left[\mathcal L_{Z_{v,\widehat v}}\!\left(
\frac{1}{3N_0}\right)\right]^{N_{\mathrm r}}.
\label{eq:scgim_average_pep}
\end{equation}
Let $\mathbf b_g(v)$ denote the bit label of candidate $v$. In terms of the
exact UPEP, the classical full-pair union bound is
\begin{equation}
P_{\mathrm b}^{\mathrm{SCGIM}}
\leq P_{\mathrm b,\mathrm{UB}}^{\mathrm{SCGIM}}
=\frac{\sum_{g,v\neq\widehat v}
d_{\mathrm H}\!\left[\mathbf b_g(v),\mathbf b_g(\widehat v)\right]
P_{\mathrm U}(v\rightarrow\widehat v)}
{B^{\mathrm{SCGIM}}K_{\mathrm t,g}P^2M}.
\label{eq:scgim_abep_union_bound}
\end{equation}
Replacing $P_{\mathrm U}(v\rightarrow\widehat v)$ in
\eqref{eq:scgim_abep_union_bound} by
\eqref{eq:scgim_average_pep} gives the tractable approximation plotted in
this work,
\begin{equation}
\widetilde P_{\mathrm b}^{\mathrm{SCGIM}}
=\frac{\sum_{g,v\neq\widehat v}
d_{\mathrm H}\!\left[\mathbf b_g(v),\mathbf b_g(\widehat v)\right]
\widetilde P(v\rightarrow\widehat v)}
{B^{\mathrm{SCGIM}}K_{\mathrm t,g}P^2M}.
\label{eq:scgim_abep_approximation}
\end{equation}

\section{Simulation Results and Discussion}
\label{sec:simulations}
This section validates the analytical results in
Sec.~\ref{sec:theoretical_analysis} and evaluates the proposed CGIM and
SCGIM through Monte Carlo simulations. Four aspects are investigated: the
accuracy of the CGIM analysis under Rayleigh fading, Nakagami-$m$ fading,
and AWGN; the BER and throughput gains over representative code-, spatial-,
and joint-domain IM benchmarks; the impact of interleaved Tx-port grouping;
and the performance--complexity tradeoff among the proposed detectors.

\subsection{BER Performance of CGIM and Code-Domain IM Scheme Benchmarks}

\begin{figure}[!t]
    \centering
    \subfloat[Rayleigh fading channel.]{%
        \includegraphics[width=0.98\columnwidth]{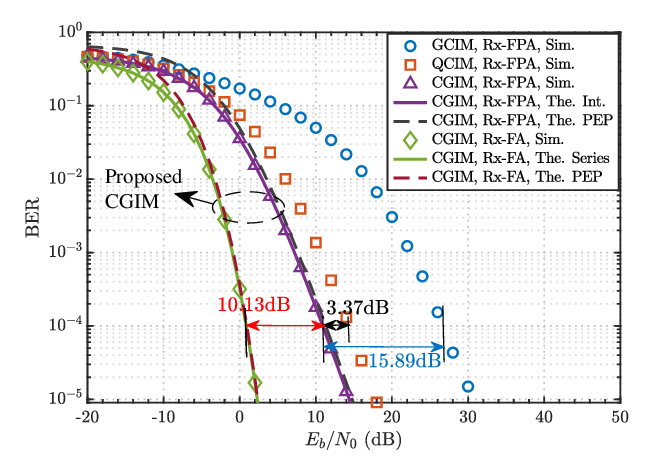}%
        \vspace{-0.12in}%
        \label{fig:cgim_rayleigh_b16}}
    \hfil
        \vspace{-0.16in}
    \subfloat[Nakagami-$m$ fading channel ($m=2$).]{%
        \includegraphics[width=0.98\columnwidth]{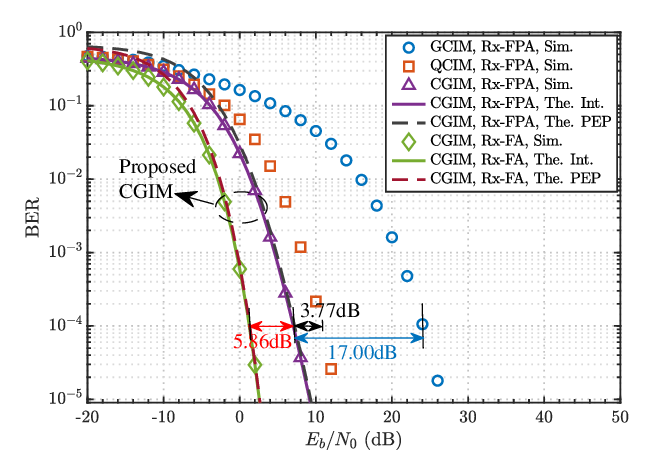}%
        \vspace{-0.12in}%
        \label{fig:cgim_nakagami_b16}}
    \hfil
     \vspace{-0.16in}
    \subfloat[AWGN channel.]{%
        \includegraphics[width=0.98\columnwidth]{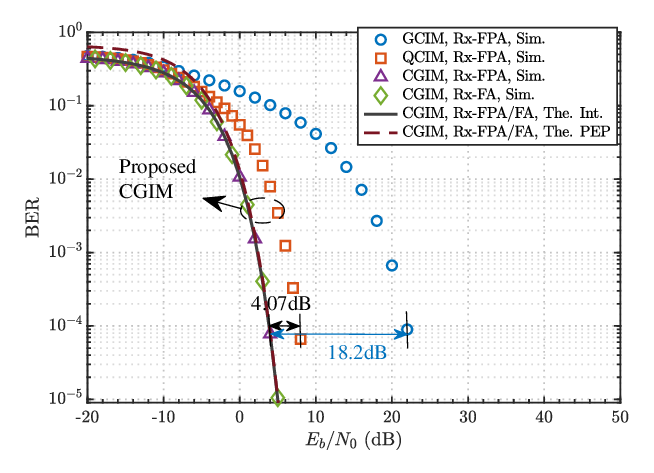}%
        \vspace{-0.12in}%
        \label{fig:cgim_awgn_b16}}
    \vspace{-0.08in}
    \caption{BER performance comparison of the proposed CGIM with existing
    code-domain index modulation schemes, including
    GCIM~\cite{Kaddoum2016GCIM} and QCIM~\cite{Cai2025QCIM}, over different
    channel models.}
    \label{fig:cgim_ber_comparison}
    \vspace{-0.08in}
\end{figure}

Fig.~\ref{fig:cgim_ber_comparison} compares the BER performance of the
proposed CGIM with GCIM~\cite{Kaddoum2016GCIM} and
QCIM~\cite{Cai2025QCIM} over Rayleigh, Nakagami-$m$, and AWGN channels. The
parameters are $L=8$, $K_r=8$, $W_{\mathrm r}=1\lambda$,
$N_{\mathrm r}=3$, and $B=16$, with $m=2$ for Nakagami-$m$ fading. Under
Rx-FPA reception and at a BER of $10^{-4}$, CGIM achieves signal-to-noise
ratio (SNR) gains of
$3.37$, $3.77$, and $4.07$ dB over QCIM in the three-type channels,
respectively; the corresponding gains over GCIM are $15.89$, $17.00$, and
$18.20$ dB. The increasing gains show that the advantage of CGIM becomes
more pronounced as channel fading weakens. Severe fading partially masks
the enlarged codeword distance enabled by the grouped code-index mapping,
whereas a more stable channel allows this distance advantage and the use of
a lower-order constellation to dominate the BER. The figure also shows that
Rx-FA provides gains of $10.13$ and $5.86$ dB over Rx-FPA in the Rayleigh
and Nakagami-$m$ channels, respectively. Selecting the port with the largest
instantaneous channel power in each receive-port set suppresses deep fades;
the stronger fluctuations of Rayleigh fading provide a larger selection
margin than Nakagami-$m$ fading with $m=2$. In AWGN, all ports experience
the same deterministic gain, so Rx-FA reduces to Rx-FPA. The analytical
curves closely agree with the simulations. The decision-region results
follow \eqref{eq:cgim_conditional_decision_events},
\eqref{eq:cgim_conditional_ber}, and \eqref{eq:cgim_average_ber}. For
Rayleigh fading, the curve labeled ``Series'' retains 16 terms per port in
\eqref{eq:cgim_marcum_series} and \eqref{eq:cgim_rxfa_gain_series}; for
Nakagami-$m$ fading, \eqref{eq:cgim_nakagami_selected_gain_cdf} is averaged
through \eqref{eq:cgim_average_ber}. For AWGN, the deterministic gain yields
\eqref{eq:cgim_awgn_ber}, leaving only the inner decision-region integral.
The ``PEP'' curves follow \eqref{eq:scgim_abep_union_bound} under
$K_{\mathrm t}=K_{\mathrm{t},g}=1$, for which SCGIM reduces to
single-fixed-transmit-antenna CGIM. To distinguish the two analyses, let
$S_\ell$ denote the ML score of candidate $\ell$. Given candidate $i$, the
exact misdecision event and its pairwise relaxation are
$\mathcal E_{i\rightarrow j}=\{S_j>\max_{\ell\ne j}S_\ell\}$ and
$\mathcal P_{i\rightarrow j}=\{S_j>S_i\}$, respectively. Since
$\mathcal E_{i\rightarrow j}\subseteq\mathcal P_{i\rightarrow j}$, PEP
evaluates a superset of the actual misdecision event. The exact events are
mutually exclusive, whereas the pairwise events may overlap; hence, the PEP
union bound is relatively loose at low SNR but becomes tight at high SNR,
where isolated nearest-neighbor errors dominate.

\subsection{SCGIM Versus Code-, Spatial-, and Joint-Domain IM Scheme Benchmarks}

\begin{figure}[!t]
    \centering
    \includegraphics[width=0.98\columnwidth]{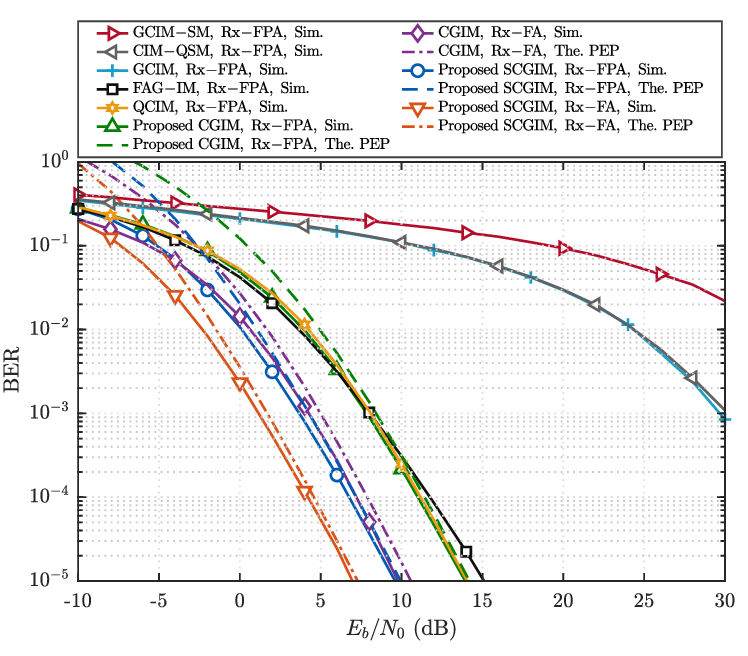}
    \vspace{-0.16in}
    \caption{BER performance comparison of the proposed SCGIM and CGIM with
    QCIM~\cite{Cai2025QCIM}, FAG-IM~\cite{Guo2025FluidAntennaIndexModulation},
    GCIM~\cite{Kaddoum2016GCIM}, GCIM-SM~\cite{Cogen2021GCIMSM}, and
    CIM-QSM~\cite{Aydin2019CIMQSM} for $B=20$ under the considered correlated
    Rayleigh channel.}
    \label{fig:scgim_full_baseline_comparison}
    \vspace{-0.10in}
\end{figure}

Fig.~\ref{fig:scgim_full_baseline_comparison} compares the BER performance of
the proposed SCGIM with representative code- and spatial-domain IM schemes over
correlated Rayleigh fading, where $B=20$, $L=16$, $K_{\mathrm t}=8$,
$W_{\mathrm t}=1\lambda$, and $N_{\mathrm r}=4$. Each receive RF chain accesses
$K_r=8$ ports distributed over $W_{\mathrm r}=1\lambda$. At a BER of $10^{-4}$, SCGIM
with Rx-FPA achieves SNR gains of approximately $4.25$, $4.42$, and $5.00$ dB
over the proposed CGIM, QCIM~\cite{Cai2025QCIM}, and the recent Tx-FA-based
FAG-IM~\cite{Guo2025FluidAntennaIndexModulation}, respectively. More
substantial improvements are observed over GCIM~\cite{Kaddoum2016GCIM},
GCIM-SM~\cite{Cogen2021GCIMSM}, and CIM-QSM~\cite{Aydin2019CIMQSM}, while
Rx-FA further improves SCGIM through receive-port selection diversity. These
results demonstrate the favorable spectral-efficiency--reliability tradeoff
of the group-wise joint indexing structure. By assigning code and Tx-port
indices within each orthogonal group, SCGIM conveys a larger fraction of the
information payload through index selection and attains $B=20$ with a lower
modulation order than the considered benchmarks, thereby reducing the burden
on the signal constellation. The interleaved Tx-port grouping further enlarges
the separation between candidate ports within each group and mitigates Tx-FAS
correlation. Hence, the observed advantage arises jointly from the group-wise
index mapping, modulation configuration, and Tx-port grouping strategy rather
than from the modulation order alone.

\begin{figure}[!t]
    \centering
    \includegraphics[width=0.98\columnwidth]{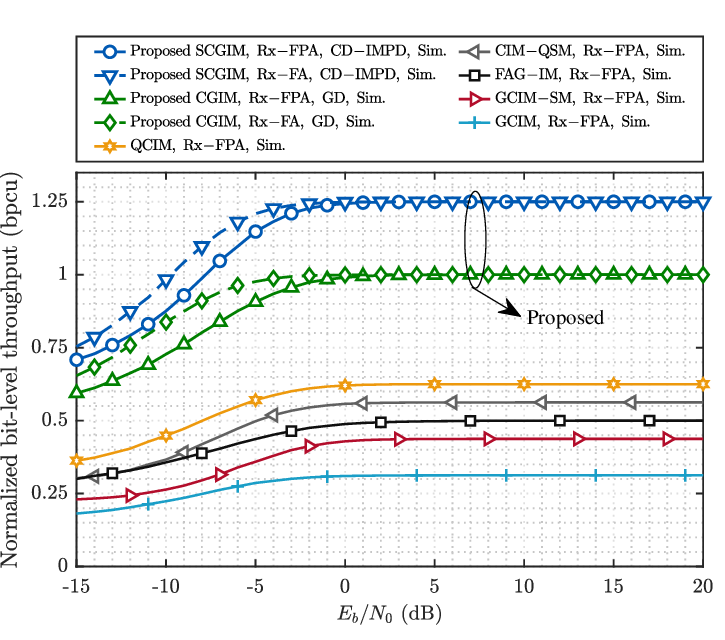}
    \vspace{-0.12in}
    \caption{Effective-throughput comparison of the proposed SCGIM and CGIM
    with code-, spatial-, and joint-domain IM benchmarks~\cite{Cai2025QCIM,
Aydin2019CIMQSM,Guo2025FluidAntennaIndexModulation,Cogen2021GCIMSM,
Kaddoum2016GCIM} under common
    code-length, modulation-order, and spatial-resource limits.}
    \label{fig:scgim_throughput_comparison}
    \vspace{-0.08in}
\end{figure}

To complement the equal-rate BER comparison, Fig.~\ref{fig:scgim_throughput_comparison}
evaluates the effective throughput under common resource limits while allowing
the payload $B_i$ to follow the indexing capability of scheme $i$. Specifically,
the available resources are limited to $L=32$, $M_{\max}=4$,
$K_{\mathrm t}=N_{\mathrm t}=16$, and $N_{\mathrm r}=4$, with $G=4$ where
applicable. The normalized bit-level throughput is defined as
$T_i=(B_i/L)(1-\mathrm{BER}_i)$~\cite{Kaddoum2016GCIM}.
Table~\ref{tab:throughput_configuration}
summarizes the resulting payloads and throughput ceilings. The proposed SCGIM
and CGIM are evaluated with both Rx-FPA and Rx-FA, whereas the benchmark
schemes employ Rx-FPA.

\begin{table}[!t]
\caption{Payloads and Maximum Effective Throughputs Under Common Resource Limits}
\label{tab:throughput_configuration}
\centering
\scriptsize
\renewcommand{\arraystretch}{1.08}
\setlength{\tabcolsep}{3.0pt}
\begin{tabular}{@{}lccc@{}}
\toprule
Scheme & Key configuration & $B_i$ & $B_i/L$ \\
\midrule
Proposed SCGIM & $G=4,\ P=8,\ K_{\mathrm{t},g}=4$ & 40 & 1.2500 \\
Proposed CGIM  & $G=4,\ P=8$                       & 32 & 1.0000 \\
QCIM           & $P=16$                            & 20 & 0.6250 \\
CIM-QSM        & $P=16,\ N_{\mathrm t}=16$         & 18 & 0.5625 \\
FAG-IM         & $G=4,\ K_{\mathrm{t},g}=4$        & 16 & 0.5000 \\
GCIM-SM        & $P=16,\ N_{\mathrm t}=16$         & 14 & 0.4375 \\
GCIM           & $P=16$                            & 10 & 0.3125 \\
\bottomrule
\end{tabular}
\vspace{-0.08in}
\end{table}

\subsection{Impact of Tx-FA Port Grouping on SCGIM}

\begin{figure}[!t]
    \centering
    \subfloat[Contiguous and interleaved Tx-port grouping.]{%
        \includegraphics[width=0.98\columnwidth]{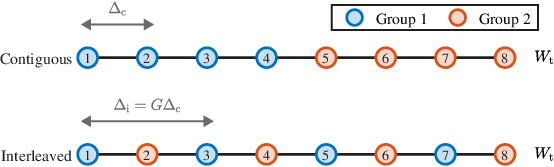}%
        \label{fig:scgim_port_grouping_schematic}}
    \par\smallskip
    \subfloat[BER comparison with CD-IMPD.]{%
        \includegraphics[width=0.98\columnwidth]{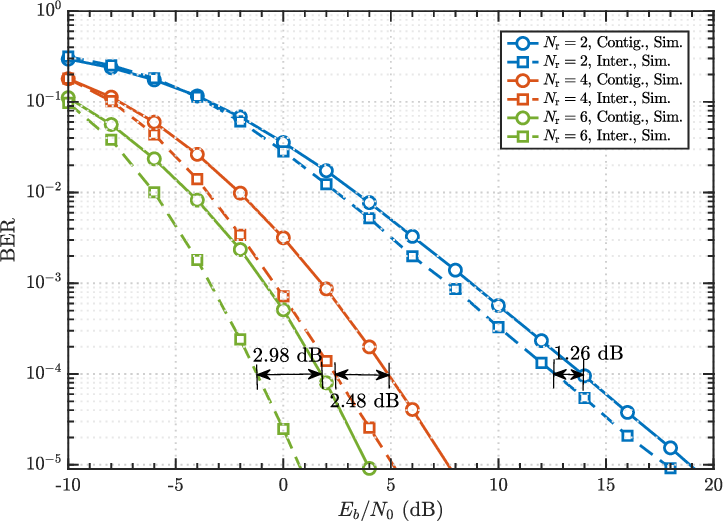}%
        \label{fig:scgim_port_grouping_ber}}
    \vspace{-0.06in}
    \caption{Contiguous and interleaved Tx-port grouping and the corresponding
    BER performance of SCGIM, where $K_{\mathrm t}=8$, $G=2$, and
    $N_{\mathrm r}\in\{2,4,6\}$.}
    \label{fig:scgim_port_grouping_comparison}
    \vspace{-0.10in}
\end{figure}

Fig.~\ref{fig:scgim_port_grouping_comparison}%
\subref{fig:scgim_port_grouping_schematic} illustrates the contiguous and
interleaved Tx-port groupings considered for $G=2$, $P=2$, $M=4$,
$K_{\mathrm t}=8$, $K_r=8$, and
$W_{\mathrm t}=W_{\mathrm r}=1\lambda$, where each group contains
$K_{\mathrm{t},g}=4$ candidate Tx ports. The receiver employs
$N_{\mathrm r}\in\{2,4,6\}$ receive RF chains and the proposed CD-IMPD detector.
Representative implementations of contiguous grouping include the block
grouping adopted in FAG-IM~\cite{Guo2025FluidAntennaIndexModulation}, which
assigns physically adjacent ports to the same group to prevent highly
correlated ports from being activated simultaneously across groups. In
SCGIM, each Tx-FA port group is instead bound to an orthogonal spreading-code
group. After group-wise despreading, different groups are separated under
ideal code orthogonality; hence, Tx-port detection is governed primarily by
the separability of candidates within the same code group. Interleaving
therefore distributes the candidates of each group over the entire Tx-FAS
aperture. If the adjacent physical-port spacing is
$\Delta_{\mathrm c}=W_{\mathrm t}\lambda/(K_{\mathrm t}-1)$, the resulting
intra-group spacing becomes $\Delta_{\mathrm i}=G\Delta_{\mathrm c}$, reducing
intra-group spatial correlation and enlarging the decision distance between
Tx-index states. Fig.~\ref{fig:scgim_port_grouping_comparison}%
\subref{fig:scgim_port_grouping_ber} compares the resulting
BERs using identical modulation parameters, channel realizations, and noise
samples, with only the Tx-port-to-code-group mapping changed. At a BER of
$10^{-4}$, interleaving provides SNR gains of approximately $1.26$, $2.48$,
and $2.98$ dB for $N_{\mathrm r}=2$, $4$, and $6$, respectively. The larger
gain with increasing $N_{\mathrm r}$ follows from the accumulation of
independent receive-branch observations, which makes the improvement in intra-group
spatial separability more pronounced. This trend is consistent with the
established observation in FA-assisted spatial IM that increasing the
physical separation between the ports associated with competing index states
mitigates spatial correlation and improves index
detection~\cite{Chen2024FAIM,Guo2025FluidAntennaIndexModulation}.

\subsection{Detection Performance and Complexity Comparison}

\begin{figure}[!t]
    \centering
    \subfloat[BER performance of the considered detectors.]{%
        \includegraphics[width=0.9\columnwidth]{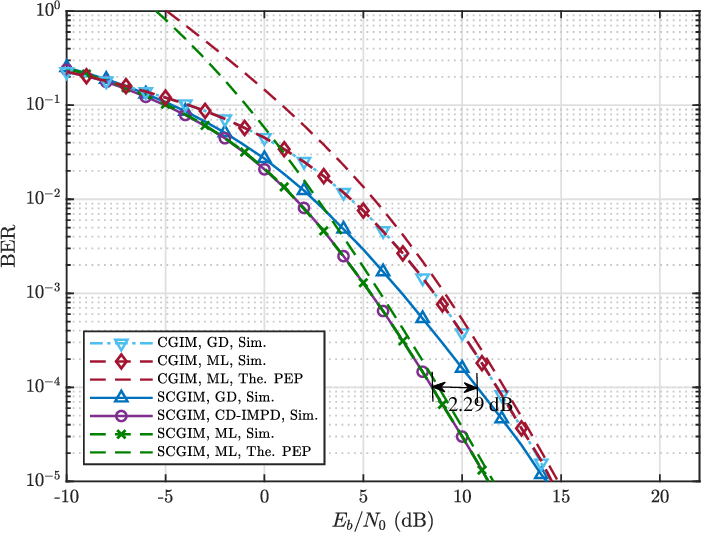}%
        \label{fig:low_complexity_ber_b32}}
    \par\medskip
    \subfloat[Detection complexity in terms of theoretical RMs and central
    processing unit (CPU) real time.]{%
        \includegraphics[width=0.98\columnwidth]{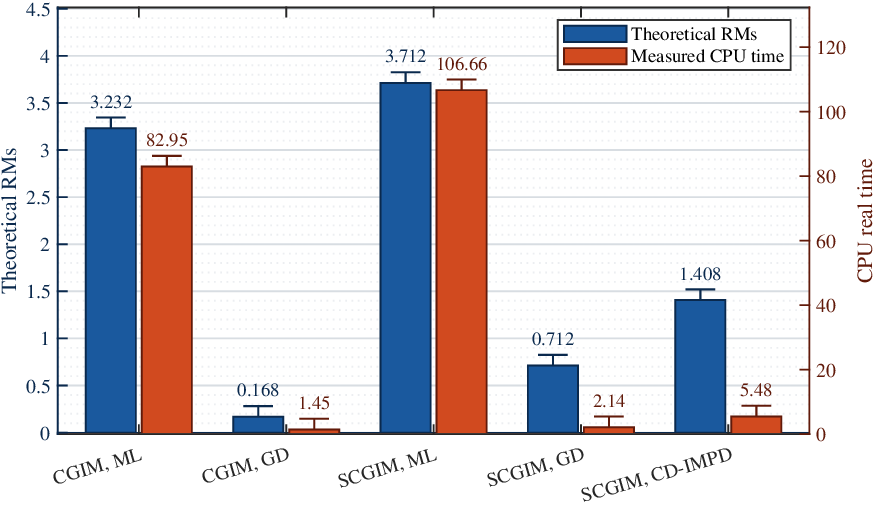}%
        \label{fig:detection_complexity_b32}}
    \vspace{-0.06in}
    \caption{BER performance and detection complexity of the proposed SCGIM
    and CGIM detectors for $B=32$.}
    \label{fig:low_complexity_performance_complexity}
\end{figure}

Fig.~\ref{fig:low_complexity_performance_complexity} jointly evaluates the
BER and detection complexity of the proposed detectors for $B=32$, $L=16$,
$G=4$, $P=2$, $K_{\mathrm t}=16$, $N_{\mathrm r}=4$, and $K_r=8$, with
$W_{\mathrm t}=W_{\mathrm r}=1\lambda$ and $K_{\mathrm{t},g}=4$. SCGIM and
CGIM employ 16-QAM and 64-QAM, respectively, to achieve the same spectral
efficiency. At a BER of $10^{-4}$, SCGIM-GD incurs a $2.29$-dB loss relative
to ML, whereas CD-IMPD and ML differ by less than $0.01$ dB. This near-ML
performance follows because
CD-IMPD preserves the coupling between the Tx-port and spreading-code
indices through soft message exchange, while GD commits to a hard port
decision before code-index detection. CGIM-GD and CGIM-ML are virtually
indistinguishable. The orthogonality of the spreading codes separates their
despread-domain observations and renders the group-wise decisions
nearly lossless. As shown in Fig.~\ref{fig:low_complexity_performance_complexity}%
\subref{fig:detection_complexity_b32},
the theoretical RMs and CPU real time of CGIM-GD are only $5.20\%$
and $1.75\%$ of those of CGIM-ML, respectively.
For SCGIM, the corresponding RM and CPU-time ratios of CD-IMPD to ML are
$37.9\%$ and $5.14\%$. Relative to SCGIM-GD, CD-IMPD requires $1.98$ times
the RMs and $2.56$ times the CPU time, and thus remains in the same
complexity order while retaining near-ML performance. The theoretical RMs and
CPU real times exhibit the same overall trend, with
GD being the least expensive, ML the most demanding, and CD-IMPD lying between
them for SCGIM. Their nonproportional numerical ratios arise from vectorized
operations, memory access, and the exponential and logarithmic evaluations
excluded from the RM count.

\section{Conclusion}
\label{sec:conclusion}

This paper developed CGIM and SCGIM for FA-assisted transceivers. CGIM partitions the spreading-code resources into parallel subsets, supporting additional code-index transmission and group-wise detection with Rx-FAS selection diversity. SCGIM further associates interleaved Tx-FA port subsets with the code subsets, extending the information-bearing dimensions to the joint spatial-code domain. ML and staged GD receivers were developed together with CD-IMPD, which retains spatial-code coupling through soft message exchange over a cycle-free factor graph. The CGIM BER was derived from the joint decision regions of its despread-domain observations under Rayleigh, Nakagami-$m$, and AWGN channels, whereas a full-pair union bound based on the Rx-FAS selected-gain density ratio and exponentially tilted quadratic-form Laplace transforms was established for SCGIM. Numerical results confirmed the analysis and demonstrated BER and throughput gains over representative IM benchmarks. They further showed that interleaved Tx-port allocation mitigates spatial correlation and that CD-IMPD achieves near-ML performance with substantially lower complexity. Overall, the proposed schemes provide an effective means of combining Tx-FA index transmission with Rx-FA selection diversity.

\bibliographystyle{IEEEtran}
\bibliography{Reference}

\end{document}